\documentclass[aps,prd,showpacs,preprintnumbers
,epsf,nofootinbib]{revtex4}
\usepackage{amsfonts,latexsym,eucal,amsmath,amssymb,times, ulem}
\usepackage[dvips]{graphicx}
\usepackage[usenames]{color}
\newcommand{\ie}{{\it i.e.}}
\newcommand{\rmx}{\mathfrak}
\definecolor{red  }{rgb}{1,0,0}
\definecolor{blue }{rgb}{0,0,1}
\definecolor{green}{rgb}{0,1,0}
\begin{document}
%\draft command makes pacs numbers print 
%\draft

\thispagestyle{empty}

%%%-----------------------------------------------------------------%%%
%%%-----------------------------------------------------------------%%%
\title{Cosmic censorship in overcharging  
a Reissner-Nordstr\"{o}m black hole via charged particle absorption}
\date{\today}
\author{Soichiro Isoyama}
\email{isoyama_at_yukawa.kyoto-u.ac.jp}
\author{Norichika Sago}
\email{sago_at_yukawa.kyoto-u.ac.jp}
\author{Takahiro Tanaka}
\email{tanaka_at_yukawa.kyoto-u.ac.jp}
\affiliation{\,\\ \,\\
Yukawa Institute for Theoretical Physics, Kyoto university,
  Kyoto, 606-8502, Japan}
%%%-----------------------------------------------------------------%%%
%%%-----------------------------------------------------------------%%%

\preprint{YITP-11-77}

%%%-----------------------------------------------------------------%%%
%%%-----------------------------------------------------------------%%%

%%%-----------------------------------------------------------------%%%
%%%-----------------------------------------------------------------%%%
\begin{abstract}
There is a claim that a static charged black hole
(Reissner-Nordstr\"{o}m black hole) can be overcharged 
by absorbing a charged test particle. 
If it is true, it might give a counter example to the weak cosmic censorship 
conjecture,
which states that spacetime singularities are never observed 
by a distant observer.
However, so far the proposed process has only been  analyzed within 
a test particle approximation. Here we claim that 
the back reaction effects 
of a charged particle cannot be neglected when judging 
whether the suggested process is really a counter example to 
the cosmic censorship conjecture or not. 
Furthermore, we argue that all the back reaction effects can be 
properly taken into account when we consider the trajectory of a 
particle on the border between the plunge and bounce orbits. 
In such marginal cases we find 
that the Reissner-Nordstr\"{o}m black hole can never be overcharged via 
the absorption of a charged particle. 
Since all the plunge orbits are expected to have a higher energy 
than the marginal orbit, we conclude that 
there is no supporting evidence that indicates 
the violation of the cosmic censorship 
in the proposed overcharging process.
\end{abstract}
%%%-----------------------------------------------------------------%%%
%%%-----------------------------------------------------------------%%%

%%%-----------------------------------------------------------------%%%
%%%-----------------------------------------------------------------%%%

\maketitle

%%%-----------------------------------------------------------------%%%
%%%-----------------------------------------------------------------%%%

%%%-----------------------------------------------------------------%%%
%\section{Introduction}  \label{Introduction}
%%%-----------------------------------------------------------------%%%
\section{Introduction}
\label{sec:Intro}
General relativity is the most successful theory of gravity 
%~\cite{Will:2005va},
and it has brought us deep a understanding of spacetime. Nevertheless,
when we evolve the Einstein equations with a well-posed initial condition, 
singularities, at which general relativity and all established theories
lose their predictability, are known to form. However, in most cases
singularities are hidden by event horizons as in the case of black holes
and cannot be seen by a distant observer. It requires extreme fine-tuning
of the initial state or unphysical equation of state to produce naked
singularities. %which can be seen from a distant observer.
This statement is known as the cosmic censorship conjecture  proposed by
Penrose~\cite{Penrose:1964wq}. Despite the tremendous efforts to prove whether
the cosmic censorship conjecture is a generic property of classical general
relativity, it still remains an open question~\cite{Wald:1997wa}. 

According to the uniqueness theorem~\cite{Mazur:1982db}, all stationary
asymptotically flat black holes in Einstein-Maxwell system are described 
by Kerr-Newman solutions, 
which are specified uniquely by the mass $M$, the charge $Q$
and the angular momentum $J$ satisfying
\begin{eqnarray}
M^2 \geq Q^2 + (J/ M)^2~.
\end{eqnarray}
When the equality is saturated, the black holes are called extremal,  
and a naked singularity appears when $M^2 < Q^2 + (J/M)^2$.   Naively,
it seems possible to form a naked singularity by throwing  matter into
a black hole to increase its charge and angular momentum. If it were
really possible, we would say that the black hole is overcharged or
overspinned via matter absorption, and it would give a counterexample
to the cosmic censorship conjecture. 

A number of previous works related to overcharging and overspinning 
a black hole support the cosmic censorship conjecture. A pioneering
work by Wald proved that neither overcharging nor overspinning is
possible when a test (charged) particle plunges into an extremal black
hole, and hence a naked singularity cannot be produced~\cite{Wald_74}.
If the particle is carrying charge or angular momentum sufficient 
to overcharge (or overspin) the black hole, it is not captured
by the black hole because of the electro-magnetic or centrifugal repulsion
force. Motivated by Wald's analysis, there have been many analyses 
attempting to supersaturate an extremal black hole by capturing a
particle and a wave packet of a classical field.~\cite{past_chargeup}. 
All these analyses indicate that such
processes at most sustain the extremal condition if the particle is
allowed to fall into the black hole. 
%%%%%%%%%%%%%%%%%%%%%%%%%%%%%%%%%%%%%%%%%%%%%%%%%%%%%%%%%%%%%%%%%%%%%%%%%%%
\footnote{
There is an exception for this statement. De Felice and Yu have pointed 
out that even an extreme RN black hole can 
turn into a naked singularity if an electrically 
neutral spinning particle is sent to an extreme RN black hole 
with appropriate initial conditions~\cite{deFelice:2001wj}.
}
%%%%%%%%%%%%%%%%%%%%%%%%%%%%%%%%%%%%%%%%%%%%%%%%%%%%%%%%%%%%%%%%%%%%%%%

Recently, however, an alternative viewpoint was raised, suggesting
the possibilities of overcharging a Reissner-Nordstr\"{o}m (RN)
black hole~\cite{Hubeny:1998ga}, 
of overspinning a Kerr black hole~\cite{Jacobson:2009kt}, 
and of overspinning or overcharging a Kerr-Newmann 
black hole~\cite{Saa:2011wq}.
The point is to consider the process of particle absorption with the initial
black hole being prepared slightly below the extremal limit.
At a first glance, these works are indicating that the violation of
cosmic censorship conjecture is possible. However, as already
mentioned in Refs.~\cite{Hubeny:1998ga} and \cite{Jacobson:2009kt}, the back
reaction effects due to the presence of a particle are not taken
into account in these analyses. In fact, it was emphasized in
Ref.~\cite{Hod:2002pm} that 
the finite size effect of a charged particle and
the contribution of the interaction energy
between the black hole and the particle can be important enough to protect
the cosmic censorship conjecture. 
In Ref.~\cite{Barausse_10}, the loss of energy due to gravitational
radiation in the spin-up process suggested in Ref.~\cite{Jacobson:2009kt}
was evaluated. Although the overspinning of a Kerr black hole can be 
avoided for some parameter choices by taking into account the effect
of this radiative loss alone, the possibility of overspinning still
remains for the other parameter choices. 

Our main purpose of this paper is to clarify the role of back reaction
effects in the gedanken experiment discussed in Ref.~\cite{Hubeny:1998ga}. 
Namely, we discuss whether a RN black hole
can be overcharged or not by the capturing of a charged particle, taking
into account all possible back reaction effects. 
The basic idea is to concentrate on the case in which the particle
is in the orbit at the border between plunge and recoil orbits,
which we call the marginal orbit. When the particle takes the
marginal orbit, it will experience an unstable equilibrium
configuration at the separatrix. Thanks to this property of the
marginal orbit, one can use an exact solution of the
Einstein-Maxwell equations to read the back reaction effects 
with sufficient accuracy for this equilibrium configuration. We will show that
the total energy of the system is always greater than the total
charge. Furthermore, using the black hole perturbation technique, 
one can evaluate the radiative loss of energy as the particle
falls into the black hole from this unstable equilibrium configuration. 
We will find that the energy loss through this process is always
negligible small. 
As a result, we will conclude that a particle in the
marginal orbit cannot overcharge the RN black hole. 
Naively, the orbits that can plunge into the black hole will have higher 
energy than the marginal orbit. Hence, our result will indicate that 
the cosmic censorship conjecture is protected even if we consider 
the process suggested in Ref.~\cite{Hubeny:1998ga}. 

The outline of this paper is as follows. 
In Sec.~\ref{sec:Overcharging} we briefly review the overcharging process 
suggested in Ref.~\cite{Hubeny:1998ga}. After explaining 
the marginal orbit in Sec.~\ref{sec:DRN}, 
we review the basic properties of an exact solution 
that corresponds to an unstable stationary equilibrium configuration 
composed of two charged sources, and we will prove that the total energy is 
always greater than the total charge for this configuration. 
In Sec.~\ref{sec:Energy_flux}, 
we will show that the radiative energy loss is always suppressed as long
as a particle falls from an equilibrium position to the black hole. 
We summarize the results in Sec.~\ref{sec:conclusion}, adding 
discussion about the implication of our results. 

In this paper, we use the units in which $G = c = 1$, 
but we explicitly write $\kappa^2 = 8 \pi G$ 
in Sec.~\ref{sec:Energy_flux} and Appendix~\ref{sec:flux}, 
respecting the original notations in Ref.~\cite{Kodama_04}.
The sign convention of the metric as $(-,+,+,+)$.
We adopt the definition of Riemann tensor and Ricci tensor given by
%\begin{eqnarray}
%$ \Gamma^{\sigma}_{\mu \nu} = (1 / 2) g^{\sigma \rho} (\partial_{\mu} 
% g_{\nu \rho} + \partial_{\nu} g_{\rho \mu} - \partial_{\rho}g_{\mu \nu})$
% and 
 $R^{\rho}_{\,\sigma \mu \nu} := \partial_{\mu} 
 \Gamma^{\rho}_{\nu \sigma} - \partial_{\nu} \Gamma^{\rho}_{\mu \sigma} 
 + \Gamma^{\rho}_{\mu \lambda} \Gamma^{\lambda}_{\nu \sigma} 
 - \Gamma^{\rho}_{\nu \lambda} \Gamma^{\lambda}_{\mu \sigma}$ 
% ~. \end{eqnarray}
%The contraction of Riemann tensor is defined as 
and 
$R_{\mu \nu} :=  R^{\alpha}_{\,\mu \alpha \nu}$. 
Fourier components of $\Psi(t)$ are defined by 
%\begin{eqnarray}
$(2 \pi)^{-1/2} \int_{-\infty}^{+\infty} 
e^{i \omega t} \Psi(t) dt$.
%-----------------------------------------------------------

%%%%%%%%%%%%%%%%%%%%%%%%%%%%%%%%%%%%%%%%%%%%%%%%%%%%%%%%%%%%%%
\section{Overcharging a Reissner-Nordstr\"{o}m black hole 
without back reaction effects}
\label{sec:Overcharging}
%%%%%%%%%%%%%%%%%%%%%%%%%%%%%%%%%%%%%%%%%%%%%%%%%%%%%%%%%%%%%%
%Before venturing into the analysis of back reaction
%effects in overcharging a RN black hole via a charged particle absorption, 
We briefly review the basic idea of overcharging a RN black hole by 
the infall of a charged particle, giving the precise meaning of 
``overcharging'', based on the discussion in Ref.~\cite{Hubeny:1998ga}.
In this section, we first focus on the test particle case in which we
neglect the back reaction effects. In this approximation we have 
a broad range of configurations whose final state exceeds the 
extremal bound. Later, we discuss how the results are modified 
by taking into account the back reaction effects.

%%%%%%%%%%%%%%%%%%%%%%%%%%%%%%%%%%%%%%%%%%%%%%%%%%%%%%%%%%%%%%
\subsection{Test particle case}
%%%%%%%%%%%%%%%%%%%%%%%%%%%%%%%%%%%%%%%%%%%%%%%%%%%%%%%%%%%%%%
We consider a point particle with mass $\mu$ and charge $q$
radially falling toward a nearly extremal RN black hole
with mass $M$ and charge $Q$. By assumption, these parameters 
satisfy $\mu < q \ll Q < M$. 
The background metric
and the vector potential of a RN black hole are given by 
\begin{eqnarray}\label{RN}
ds^2  &\!\! =&\!\! - f(r) dt^2 + \frac{1}{f(r)} dr^2 + 
r^2 \left(d \theta^2 + \rm{sin}^2 \theta d \phi^2 \right)~, \cr
A_{t} &\!\!=&\!\! -\frac{Q}{r}, \qquad A_{r} = A_{\theta} = A_{\phi} = 0~,
\end{eqnarray}
where
\begin{equation}
 f(r) := 1 - \frac{2M}{r} + \frac{Q^2}{r^2}~.
\label{RNf}
\end{equation}

Due to the spherical symmetry of a RN black hole, we can choose the 
coordinate system such that the trajectory of a particle 
is along the axis without loss of generality.
Then, the trajectory is represented by
$z^\alpha(s) = ( T(s), R(s), 0, 0)$ with the proper time $s$ along
the world line of the particle.  
The equations of motion of the particle are given by 
\begin{eqnarray}
\label{dR-ds}
\left( \frac{dR}{ds} \right)^2 &\!\! =&\!\!
\frac{1}{\mu^2} 
\left(E-{qQ\over R}\right)^2-f(R)=: V^{(\rm{o})}(R)~,\\
%( E - V_{-}^{(\rm{o})} (R) ) ( E - V_{+}^{(\rm{o})} (R) )
\left( \frac{dT}{ds} \right) &\!\! =&\!\!
\frac{1}{\mu f(R)} \left( E - \frac{qQ}{R} \right).
\label{dT-ds}
\end{eqnarray}
The energy of the particle $E$ is defined by 
$E := - (\partial_{t})^a (\mu u_a + q A_{a})$ where 
$(\partial_{t})$ is the Killing field associated with 
the time coordinate $t$ 
and $u^a := d z^a /ds$ is the four velocity of the particle. 
Using the $t$-component of the equations of motion~(\ref{dT-ds}),
it is shown that $E$ is constant along the orbit.

Radial orbits toward the black hole can be classified into 
three classes: plunge, bounce and marginal
orbits. In the case of plunge orbits, a particle falls into
the black hole horizon without changing the
direction of motion. 
In the case of bounce orbits, there exists
a turning point outside the horizon, at which $(dR/ds)^2=0$.
The particle is reflected back at this point due to the electric 
repulsion force. The marginal orbit is on a separatrix
between plunge and bounce orbits. In this case, the particle 
gradually approaches an unstable equilibrium position at $r=r_{0}$. 
The position of $r_{0}$ and the energy of the marginal orbit 
%$E_{\rm M}$ 
are simultaneously determined by the conditions 
$dR/ds = d^2R/ds^2 = 0$.

%The basic concept to overcharge a RN black hole is  sending a
%charged particle into the RN black hole and  makes the black hole
%jump over the extreme limit. More precisely, this concept is
%formulated as 
In the above setup, 
if the following two conditions are satisfied, we would say 
that a RN black hole is possibly overcharged. 
The first condition is that the particle is in a plunge orbit.
For plunge orbits there is no turning point where the radial 
velocity becomes zero. 
Therefore the condition is 
\begin{equation}
\label{falling-cond}
\left(\frac{dR}{ds} \right)^2=V^{(\rm{o})}(R)  > 0,\qquad 
{\rm for}\quad R \geq r_+~,\quad 
\mbox{(absorption condition)}
\end{equation}	
where $r_+ := M + \sqrt{M^2 - Q^2}$ is the radial coordinate 
of the event horizon.
The second condition is on the total energy of the final state. 
After the absorption of the particle, the system will approach
another RN geometry with mass $M_{\rm final}=M+E$ and charge
$Q_{\rm final}=Q+q$. The condition that the final 
RN geometry exceeds the extremal bound is given by 
\begin{equation}
\label{extremality}
M + E < Q + q~. \qquad 
\mbox{(overcharging condition)}
\end{equation}

In Ref.~\cite{Hubeny:1998ga}, it was demonstrated that
radial orbits in a rather wide range of parameter space 
satisfy the conditions (\ref{falling-cond}) and (\ref{extremality}). 
In fact,
both the absorption and overcharging conditions are satisfied 
for  
\begin{equation}
 1 < b < a~, \quad c < \sqrt{a^2 - b^2}~, 
\end{equation}
under the parametrization 
\begin{eqnarray}
\label{e-parametrization-particle+RN}
M   &\!\! :=&\!\! 1 + 2 \epsilon^2~, \qquad Q   := 1~, \qquad
E   := a \epsilon - 2 b \epsilon^2~, \qquad
q   := a \epsilon~, \qquad
\mu := c \epsilon~, 
\end{eqnarray}
%~~~~~~~~~~~~~~~~~~~~~~~~~~~~~~~~~~~~~~~~~~~~~~~~~~~~~~~~~~~~~
where $0< \epsilon \ll 1$  and the coefficients
$a,~b$ and $c$ are assumed to be $O(1)$ real numbers.
Here the overcharging is discussed at the level of $O(\epsilon^2)$, 
while the mass and the charge of a point particle are $O(\epsilon)$. 
Although the back reaction 
effects are completely neglected, the analysis in Ref.~\cite{Hubeny:1998ga} 
is in sharp contrast with previous works~\cite{past_chargeup}. 
The previous analyses are restricted to $O(\epsilon)$ 
and only the extremal limit is considered for the initial black hole. 

%%%%%%%%%%%%%%%%%%%%%%%%%%%%%%%%%%%%%%%%%%%%%%%%%%%%%%%%%%%%%%
\subsection{Back reaction effects on the overcharging process}
\label{subsec:Backreaction}
%%%%%%%%%%%%%%%%%%%%%%%%%%%%%%%%%%%%%%%%%%%%%%%%%%%%%%%%%%%%%%
As was already emphasized in Ref.~\cite{Hubeny:1998ga}, 
the analysis in the previous
section is not sufficient to conclude that a nearly 
extremal RN black hole can be overcharged through 
the absorption of a charged particle. 
We have to compute the total amount of energy in the final state of the 
whole system to the accuracy of $O(\epsilon^2)$, 
taking into account the effects of back reaction~\cite{Johnston,Hod:2002pm}.
If the evaluation to this accuracy is accomplished, one can judge whether the 
overcharging condition~\eqref{extremality} is satisfied or not. 
At the same time, we have to examine how the absorption 
condition~\eqref{falling-cond} is modified once the 
back reaction effects are taken into account.

To achieve this goal, 
we need to calculate both the particle motion with the 
self-force and the self-energy of the system to $O(\epsilon^2)$. 
However, these quantities have never been computed to such a high order 
even in the case of Schwarzschild background.
Here, we propose to bypass these difficult tasks by 
focusing on the marginal orbits that pass through an unstable equilibrium
state. 
The basic strategy that we use here is to relate the unstable
equilibrium state to an exact solution known as the 
double Reissner-Nordstr\"{o}m static 
solution~\cite{Alekseev_07}
(See also Ref.~\cite{Manko:2007hi})
%%%%%------------------------------------------------------
\footnote{
Within the linear perturbation, the global perturbative solution 
describing the equilibrium configuration of the RN black hole and 
a charged particle was constructed in Ref.~\cite{Bini:2006dp}.}. 
Since the latter is an exact solution, 
it includes all the backreaction effects. Therefore, we can read the
total energy of the system in the unstable equilibrium state 
from this solution. 

Then, we compute the energy emitted to infinity through 
the infall of the particle from the equilibrium
state to the black hole horizon to the accuracy of $O(\epsilon^2)$, which 
can be readily achieved using 
the standard black hole perturbation method. 
To compute the energy flux emitted to infinity,
back reaction effects on the trajectory of a particle can be neglected 
since they give only higher order corrections to the energy flux. 
As a result, we obtain a sufficiently accurate estimate for 
the total energy of the final state for the marginally plunging 
orbits. 
In the two successive sections, we prove that 
the overcharging is impossible for any of such marginal orbits. 

To extend our argument to more general cases, 
we assume that the other plunge orbits result in 
final states that have higher energies than 
the final state in the case of the marginal trajectory, 
whose energy is deduced under the conditions that the mass and charge of 
the initial black hole and the charge of the plunging particle are fixed. 
Once we accept this rather reasonable assumption, the avoidance 
of overcharging in the marginal cases is extended to general 
plunge orbits. Although, strictly speaking, 
we cannot verify this last statement, 
we can at least claim that there is no evidence that supports the 
possibility of violating the weak 
cosmic censorship conjecture in the present context.

%%%%%%%%%%%%%%%%%%%%%%%%%%%%%%%%%%%%%%%%%%%%%%%%%%%%%%%%%%%%%%%   Sec 
\section{Total energy of the system with 
a charged black hole and a charged particle
in equilibrium}
\label{sec:DRN}
%%%%%%%%%%%%%%%%%%%%%%%%%%%%%%%%%%%%%%%%%%%%%%%%%%%%%%%%%%%%%%%   Sec 
As a first step to examine the overcharging condition~\eqref{extremality} 
for the marginal orbits, we here extract the total energy of the system 
composed of a charged black hole and a charged particle 
in the equilibrium state, $E_{\rm eq}$, using the double Reissner-Nordstr\"{o}m
(DRN) solution of 
the Einstein-Maxwell equations~\cite{Alekseev_07, Manko:2007hi}. 
On the other hand, the total charge is simply given by the sum of the charges 
assingned to respective objects as $Q_{\rm total} = q + Q$ since the electric 
charge is not carried by a field.
In this section, we will show that the energy at the equilibrium state
is always larger than the total charge, $E_{\rm eq}-(q+Q)=O(\epsilon^2)>0$.

%%%%%%%%%%%%%%%%%%%%%%%%%%%%%%%%%%%%%%%%%%%%%%%%%%%%%%%%%%%%%%%%%%%%%%%%
\subsection{The double Reissner-Nordstr\"{o}m solution}
\label{subsec:DRN}
%%%%%%%%%%%%%%%%%%%%%%%%%%%%%%%%%%%%%%%%%%%%%%%%%%%%%%%%%%%%%%%%%%%%%%%%
We use the form of the DRN solution 
presented in Ref.~\cite{Alekseev_07} 
in the same notation, except for the signature of the metric, $(-,+,+,+)$.
The DRN solution is characterized by five parameters: masses of
the RN sources $m_1,~m_2$, their charges $e_1,~e_2$, and the separation
between them $\ell$. 
It should be emphasized that the mass parameters $m_1$ and $m_2$ do not 
refer to the rest masses of the RN sources~\cite{Alekseev_07}. 
As we explain later, there is one relation among 
these five parameters, and hence four of them are independent.  
With these parameters, the metric and the vector 
 potential are written in the cylindrical Weyl coordinates as 
\begin{eqnarray}
&& ds^2 = - H(\rho,~z) dt^2 + F(\rho,~z) (d \rho^2 + dz^2) 
+ \frac{\rho^2}{H(\rho,~z)} d\phi^2~,
\label{double_RN_metric} \\
&&\qquad A_{t} = - \Phi(\rho,~z), \qquad A_{\rho} = A_{z} = A_{\phi} = 0~.
\label{double_RN_EM}
\end{eqnarray}
To express the functions $H$, $F$ and $\Phi$ in a simple form, 
we introduce two sets of bipolar coordinates ($r_1,\theta_1$) and
($r_2,\theta_2$) defined by
\begin{eqnarray}\label{bipolar}
 \begin{cases}
 \rho = \sqrt{(r_{1} - m_{1})^2 - \sigma_1^2} ~{\rm{sin}} \theta_1~, \\
 z   = ~z_1 + (r_1 - m_1) ~{\rm{cos}} \theta_1~,
 \end{cases}
 \begin{cases}
 \rho = \sqrt{(r_2 - m_2)^2 - \sigma_2^2} ~{\rm{sin}} \theta_2~, \\
 z   = ~z_2 + (r_2 - m_2) ~{\rm{cos}} \theta_2~,
 \end{cases}
\end{eqnarray}
where $z_1$ and $z_2$ are set to satisfy $\ell=z_2-z_1>0$, and 
$\sigma_1^2$ and $\sigma_2^2$ are defined by 
\begin{eqnarray}
\label{def-sigma-double-RN}
\sigma_1^2 = m_1^2 - e_1^2 + 2 e_1 \gamma~, \quad 
\sigma_2^2 = m_2^2 - e_2^2 - 2 e_2 \gamma~. \quad 
\end{eqnarray}
Using these coordinates, the functions $H,~F$ and $\Phi$ are written as
\begin{eqnarray}
H(\rho, z) &\!\! =&\!\! \left[ (r_1 -m_1)^2 -\sigma_1^2 + \gamma^2 {\rm{sin}^2} \theta_2 \right]
 \left[ (r_2 -m_2)^2 -\sigma_2^2 + \gamma^2 {\rm{sin}^2} \theta_1 \right] 
 \mathcal{D}^{-2}~,
\label{double_RN_H} \\
F(\rho, z) &\!\! =&\!\! \left[ (r_1 -m_1)^2 -\sigma_1^2 \cos^2\theta_1 \right]^{-1}
\left[ (r_2 -m_2)^2 -\sigma_2^2 \cos^2\theta_2 \right]^{-1}  \mathcal{D}^2~,
\label{double_RN_f} \\
\Phi(\rho, z) &\!\!=&\!\! \left[ (e_1 - \gamma)(r_2 - m_2) + (e_2 + \gamma)(r_1 - m_1) 
+ \gamma (m_1{\rm{cos}} \theta_1 + m_2{\rm{cos}} \theta_2) \right] 
~\mathcal{D}^{-1}~,
\label{double_RN_Phi}
\end{eqnarray}
where
\begin{equation}
\mathcal{D} = 
r_1 r_2  - (e_1 -\gamma -\gamma ~{\rm{cos}} \theta_2)
(e_2 + \gamma -\gamma ~{\rm{cos}} \theta_1)~,
\label{double_RN_D}
\end{equation}
and the parameter $\gamma$ is defined by
\begin{eqnarray}
\label{def-gamma-double-RN}
\gamma = (m_2 e_1 - m_1 e_2) (\ell + m_1 + m_2)^{-1}~.
\end{eqnarray}
The parameter $\sigma^2_1$ is chosen to be negative 
while $\sigma^2_2$ positive. 
The object labeled with 1 is 
a naked singularity that has a critical spheroid at 
$\{ r_1 = m_1,~0 \leq \rho \leq |\sigma_1|,~z = z_1 \}$, 
while the object labeled with 2 is a black hole
surrounded by an event horizon at
$\{ \rho = 0,~z_2 - \sigma_2 \leq z \leq z_2 + \sigma_2 \}$.  
The position of the naked singularity is in the region $r_1< m_1$, 
which is not covered by the original coordinates $(\rho, z)$. 
To guarantee the existence of an equilibrium state without any conical
singularities between the two objects, 
a balance condition,
\begin{eqnarray}\label{balance-eq}
m_1 m_2 = (e_1 - \gamma) (e_2 + \gamma)~, 
\end{eqnarray}
must be satisfied. This gives one relation among five parameters 
$(m_1, m_2, e_1, e_2, \ell)$, which can be 
satisfied only when $\sigma_1^2\sigma_2^2\leq 0$. 
Namely, the DRN solution should consist either of 
a black hole and a naked singularity, or 
of two extremal black holes. 
The formar case requires that 
$\sigma^2_1$ and $\sigma_2^2$ should have opposite signs, as we have
chosen above.
The latter is a special 
case of the Majumdar-Papapetrou solution~\cite{MP}. 

The total mass and charge of the DRN solution can be read from the
asymptotic forms of the metric functions (\ref{double_RN_H})
and (\ref{double_RN_f}), and the electric potential
(\ref{double_RN_Phi}).  
At the space-like infinity, defined by 
the limit $\rho \to \infty$ or $z \to \infty$, the two bipolar
coordinates $(r_1, \theta_1)$ and $(r_2, \theta_2)$ coincide with
the spherical coordinates $(r, \theta)$ defined by
$\rho = r {\rm{sin}} \theta$ and $z = r {\rm{cos}} \theta$.
Thus, 
the metric functions and the electric potential are expanded as:
\begin{eqnarray}\label{double-RN-asymptotic}
H      &\!\! =&\!\! 
1 - \frac{m_1 + m_2}{r} + O \left( \frac{1}{r^2} \right)~, \cr
F^{-1} &\!\! =&\!\! 
1 - \frac{m_1 + m_2}{r} + O \left( \frac{1}{r^2} \right)~, \cr
\Phi   &\!\! =&\!\!     \frac{e_1 + e_2}{r} + O \left( \frac{1}{r^2} \right)~,
\end{eqnarray}
in this region.
For asymptotically flat spacetime, the coefficients
of the $O(1/r)$ terms of the metric function and the electric potential
correspond to the total energy and charge of the system, respectively.
Thus we find that the total energy is $m_1 + m_2$ and the total charge is 
$e_1 + e_2$. 

%%%%%%%%%%%%%%%%%%%%%%%%%%%%%%%%%%%%%%%%%%%%%%%%%%%%%%%%%%%%%%%%%%%%%%%%
\subsection{Mass dominance of the DRN solution}
\label{subsec:energetics-DRN}
%%%%%%%%%%%%%%%%%%%%%%%%%%%%%%%%%%%%%%%%%%%%%%%%%%%%%%%%%%%%%%%%%%%%%%%%
In this subsection, we study the DRN solution 
consisting of a black hole 
with charge $e_2$ and mass $m_2$ satisfying 
$\sigma_2^2 > 0$, 
and a naked singularity 
with charge $e_1$ and mass $m_1$ satisfying 
$\sigma_1^2 < 0$.  We assume
\begin{equation}\label{DRN-cond}
m_2 > m_1 > 0~, 
\qquad
e_2 > ~e_1 > 0~, 
\end{equation}
which means that the mass and the charge of the black hole 
are larger than those of the naked singularity. 
Under these conditions, we show that the total mass is larger
than the total charge, \ie ~$m_1+m_2>e_1+e_2$.

To this end, we re-express the balance equation (\ref{balance-eq}) as 
\begin{equation}
I(\gamma) := \gamma^2 + (e_2 - e_1) \gamma + m_1 m_2 - e_1 e_2=0~.
\label{balance-func}
\end{equation}
In addition, from $\sigma_1^2<0$, $\sigma_2^2>0$, 
and using Eq.~\eqref{def-sigma-double-RN},   
we obtain the conditions for the parameter $\gamma$ as
\begin{eqnarray}\label{g-range1}
\gamma < \frac{e_1^2 - m_1^2}{2 e_1}~,
\quad
\gamma < \frac{m_2^2 - e_2^2}{2 e_2}~,
\end{eqnarray}
and, from the positivity of the 
separation $\ell > 0$ and using Eq.~(\ref{def-gamma-double-RN}), 
we find 
\begin{eqnarray}\label{g-range2}
 \begin{cases}
 \displaystyle \frac{m_2 e_1 - m_1 e_2}{m_1 + m_2} < \gamma < 0,
 & \mbox{for}~~
   m_2 e_1 - m_1 e_2 < 0, \\
 & \\
 0< \gamma < \displaystyle \frac{m_2 e_1 - m_1 e_2}{m_1 + m_2},
 & \mbox{for}~~
   m_2 e_1 - m_1 e_2 > 0. 
 \end{cases}
\end{eqnarray}
In this context, the balance condition (\ref{balance-func})
is interpreted as the condition that
the quadratic equation of $\gamma$ has a solution in the range 
specified by the conditions, (\ref{g-range1}) and (\ref{g-range2}).

Now we examine if a solution exists for three cases:
(i) $\gamma<0$ and $m_1 m_2 > e_1 e_2$,
(ii) $\gamma<0$ and $m_1 m_2 < e_1 e_2$, and
(iii) $\gamma>0$ and $m_1 m_2 < e_1 e_2$.  
For $\gamma>0$, the case with $m_1 m_2 > e_1 e_2$ is immediately 
excluded by the condition (\ref{balance-func}).\newline
Case (i): 
In this case, the conditions~\eqref{g-range1} and \eqref{g-range2} 
imply 
\begin{eqnarray}
\label{g-range-temp}
-\gamma > \frac{m_1^2 - e_1^2}{2 e_1}~,
\quad
-\gamma > \frac{e_2^2 - m_2^2}{2 e_2}~,
\quad
-\gamma < \frac{ m_1 e_2 - m_2 e_1 }{m_1 + m_2}~.
\end{eqnarray}
%Given that $m_2 > m_1 > 0$ implies \&$m_1 / (m_1 + m_2) > 0$ and $m_2 / (m_1 + m_2) > 0$, 
The first two inequalities in~\eqref{g-range-temp} lead to
\begin{eqnarray}\label{g-ineq-1}
-\gamma &\!\! >&\!\! \frac{1}{m_1 + m_2} 
\left[
\frac{m_2 (m_1^2 - e_1^2)}{2 e_1} + \frac{m_1 (e_2^2 - m_2^2)}{2 e_2}
\right] \cr
&\!\! =&\!\!  \frac{m_1 e_2 - m_2 e_1}{2 ( m_1 + m_2) } 
\left[  1 + \frac{m_1 m_2}{e_1 e_2} \right] \cr
&\!\! >&\!\! \frac{m_1 e_2 - m_2 e_1}{ m_1 + m_2 }~,
\end{eqnarray}
where we used $m_1 m_2 > e_1 e_2$ in the last line.
However, the inequality~\eqref{g-ineq-1} is in contradiction with the last 
condition in~\eqref{g-range-temp}.
Thus, case (i) is excluded.\newline
Case (ii): In this case, the balance condition (\ref{balance-func}) 
implies 
\begin{equation}
\gamma^2 + (e_2-e_1)\gamma = e_1 e_2 - m_1 m_2 > 0,
\end{equation}
which leads to the inequality
\begin{equation}
 \gamma+ (e_2 - e_1) < 0~.
\end{equation}
Combining this inequality with Eqs.~\eqref{g-range2}, we find
\begin{eqnarray}
\frac{m_1 e_2 - m_2 e_1}{ m_1 + m_2 } > -\gamma > e_2 - e_1~,
\end{eqnarray}
and hence we have 
\begin{eqnarray}
m_1 e_1 > m_2 e_2~.
\end{eqnarray}
However, the last inequality $m_1 e_1 > m_2 e_2$ is in contradiction 
with the assumption $m_2 > m_1 > 0$ and $e_2 > e_1 > 0$. 
Thus, case (ii) is also excluded. \newline
Case (iii):
Recalling that $I(0) = m_1 m_2 - e_1 e_2 < 0$, the necessary condition 
that $I(\gamma) = 0$ has a 
positive root satisfying the condition~(\ref{g-range2}) is
\begin{eqnarray}
\label{I(g)-conditions}
I \left( \frac{m_2 e_1 - m_1 e_2}{m_1 + m_2} \right)
= \frac{m_1 m_2}{(m_1 + m_2)^2} 
\left[ (m_1 + m_2)^2 - (e_1 + e_2)^2 \right] > 0~.
\end{eqnarray}
This inequality leads to
\begin{equation}
\label{DRN-inequality}
m_1 + m_2 > e_1 + e_2~.
\end{equation}

Since both case (i) and case (ii)
are not allowed, the only possibility is case (iii), and 
we also find that $\gamma$ cannot be negative simultaneously. 
In case (iii) the inequality (\ref{DRN-inequality}) was proven. 
Hence, to conclude, under the assumptions $m_2 > m_1 > 0$ and $e_2 > e_1 > 0$,
we have shown that 
\begin{equation}
\label{extremality:balance}
E_{\rm{eq}} > q + Q,
\end{equation}
which means that the total energy never goes below the extremal bound 
in the equilibrium configuration.

%---------------------------------------------------------
%%%%%%%%%%%%%%%%%%%%%%%%%%%%%%%%%%%%%%%%%%%%%%%%%%%%%%%%%%%%%%%%%%%%%%%%
\subsection{Remarks on the use of the DRN solution}
\label{subsec:Remark-DRN}
%%%%%%%%%%%%%%%%%%%%%%%%%%%%%%%%%%%%%%%%%%%%%%%%%%%%%%%%%%%%%%%%%%%%%%%%
In the preceding subsection, we used the DRN solution as such 
that describes an equilibrium state of 
a charged particle in a black hole spacetime.
However, strictly speaking, we also have to prove that the charged singularity
in the DRN solution is a valid approximation for a charged particle. 
An arbitrary solution that allows a singularity may in general possess higher 
multipole moments other than the monopole. In that case 
the charged particle will have an extra self-energy due 
to higher multipole distortion. Thus, one may suspect that 
there might be another equilibrium configuration that has 
lower total energy with the same total charge. 
%We can understand even intuitively that this possibility is 
%unlikely. The form 
%of this DRN solution is symmetric under the interchange of 
%points 1 and 2. Namely, the singular point particle is 
%described in the same functional form as the black hole. 
%Since we know that the black hole cannot possess any 
%extra multipole moments, 
%it is natural to anticipate that so is 
%this naked singularity, which is 
%different from the balck hole by the choice 
%of parameters. 
Hence, it would be necessary to explicitly 
demonstrate that the charged particle in the DRN solution does not possess 
higher order multipole moments that are large enough to contribute 
to the total energy of $O(\epsilon^2)$. 
A detailed discussion is given in Appendix~\ref{sec:singularity}, 
and the outline is as follows. 
We first show that the behavior near the singularity in 
the DRN solution can be described by a perturbed RN configuration. 
Perturbative expansion does not break down even near the 
singularity in the DRN solution. 
From this expression we can read the maximum amplitude of 
the multipole moments that the particle possesses. The largest 
component is the dipole and its magnitude is $O(\epsilon^3)$, 
and we find that the contribution of such a dipole moment to 
the total energy is at most $O(\epsilon^3)$. 

%%%%%%%%%%%%%%%%%%%%%%%%%%%%%%%%%%%%%%%%%%%%%%%%%%%%%%%%%%%%%%%   Sec 
\section{Energy flux radiated from a charged particle}
\label{sec:Energy_flux}
%%%%%%%%%%%%%%%%%%%%%%%%%%%%%%%%%%%%%%%%%%%%%%%%%%%%%%%%%%%%%%%   Sec 
In the preceding section, with the help of the DRN solution, 
we showed that the total energy of the equilibrium
state of a charged particle in a RN black hole 
cannot be less or equal to its total charge.
This suggests that the system in equilibrium does not exceed 
the extremal bound,
but it is not sufficient yet to exclude the possibility that a RN
black hole is overcharged by absorbing a charged particle.
The effects of the electromagnetic and gravitational radiation can
reduce the total energy as the particle falls from the 
equilibrium position to the event horizon of the black hole.
Then the total energy in the final state will be given by 
$E_{\rm{eq}} -  E_{\infty}$ with $E_{\infty}$ being 
the energy radiated away to infinity. 
In this section, we estimate $E_{\infty}$ using linear perturbation theory. 
We examine if the signature of $E_{\rm eq}- q - Q - E_{\infty}$ can be 
negative, to see whether the overcharging in the final 
configuration is possible or not.

%%%%%%%%%%%%%%%%%%%%%%%%%%%%%%%%%%%%%%%%%%%%%%%%%%%%%%%%%%%%%%
\subsection{Correspondence between two different pictures}
\label{subsec:BHPvsDRN}
%%%%%%%%%%%%%%%%%%%%%%%%%%%%%%%%%%%%%%%%%%%%%%%%%%%%%%%%%%
Before evaluating the radiation energy emitted to infinity, 
we would like to briefly mention the correspondence between the DRN solution 
and the equilibrium configuration of a test particle in a RN black hole. 
Here we refer to the latter as the black hole perturbation (BHP) picture. 
Our current interest is in the case in which the charge of the test particle 
is $O(\epsilon)$, while the difference between the total energy 
and the total charge is $O(\epsilon^2)$:  
\begin{equation}\label{trial}
e_1 := a \epsilon~,~~~e_2 := 1~,~~~m_1 := O(\epsilon)~,~~~
m_2 := 1 + \varpi (1 + O(\epsilon))~,~~~
(m_1 + m_2) - (e_1 + e_2) = O(\epsilon^2)~. 
\end{equation}
As a starting point we do not exclude  
the possibility that 
$\varpi$ is as large as $O(\epsilon)$. 
There might be more subtle cases if we consider the possibility that
the difference $(m_1 + m_2) - (e_1 + e_2)$ is $O(\epsilon^3)$ or higher, 
but such cases are beyond the scope of the present paper. 

Then, as shown in Appendix~\ref{sec:e-prameter}, 
the parameters used to describe the DRN solution are expanded in terms of 
$\epsilon$ as 
\begin{equation}
 e_1    =  a \epsilon~,\quad
 e_2    =  1~,\quad
 \gamma =  g \epsilon^2,\quad
 \ell \gamma   =  \tilde g \epsilon^3, 
\label{e-parameter2}
\end{equation}
where $a$ and $\tilde g$ are expected to be $O(1)$, while $g$ can be 
either $O(1)$ or even smaller.  
If $g$ is of higher order in $\epsilon$, $\ell$ should be large 
so as to maintain $\ell \gamma=O(\epsilon^3)$. 
The meaning of $a$ is obviously identical to that in the BHP picture. 
Substituting Eqs.~(\ref{e-parameter2}) into 
Eqs.~(\ref{def-sigma-double-RN}), 
\eqref{def-gamma-double-RN} and \eqref{balance-eq}, 
we obtain approximations for the mass parameters $(m_1,m_2)$
and $(\sigma_1,\sigma_2)$ as
\begin{eqnarray}
 m_1   = a \epsilon - g \epsilon^2-{\tilde g\over 2}\epsilon^3~,\quad
 m_2   = 1 + \left( g+ \frac{\tilde g}{2a} \right)\epsilon^2~, \quad
 \sigma_1^2   = ( g^2 - a \tilde g)\epsilon^4~,\quad
 \sigma_2^2   = \frac{ \tilde g}{a}\epsilon^2~, 
\label{expanded-variables-DRN}
\end{eqnarray}
where the terms of higher order in $\epsilon$ are neglected. 
The relations~\eqref{expanded-variables-DRN} enable us to evaluate
the difference between the total energy and the total charge as
\begin{equation}
m_1 + m_2 - (e_1 + e_2)
= \frac{\tilde g}{2a}\epsilon^2 + O(\epsilon^3), 
\label{eq:difference-EC}
\end{equation}
which is positive as long as $\gamma$ and $\ell$ are 
positive. This is, of course, consistent with the more general argument 
given in the preceding section. 

The relation between the parameters $\tilde g$ and $g$ in the 
DRN solution and the parameters in the BHP picture 
is examined in Appendix~\ref{subsec:Mapping}, and is given by 
\begin{eqnarray}
\label{mapping}
 \tilde g=4 a+O(\epsilon)
~,\qquad
 g  =  2 \sqrt{a^2 - c^2}+O(\epsilon)=2 b+O(\epsilon).
\end{eqnarray}
(Here $b$ is simply determined by the equilibrium
condition~\eqref{discriminant}.) From these expressions we can confirm 
that there are counterparts of $g$ and $\tilde g$ 
for any values of $a$ and $c$. 
Using the above relation, we finally obtain 
\begin{equation}
m_1 + m_2 - (e_1 + e_2)
= 2\epsilon^2 + O(\epsilon^3).  
\end{equation}

In the analysis neglecting all back reaction effects, as in 
Ref.~\cite{Hubeny:1998ga},
our interest was restricted to the parameter region 
with $b>1$ that allows overcharging in the final state.
Under this restriction, $g$ must be definitely $O(1)$. 
As a result, $\ell$ must be $O(\epsilon)$. 
As shown in Appendix.~\ref{subsec:Mapping}, 
the corresponding proper separation is O(1). 
However, now we know that, once all the back reaction effects 
are taken into account, the difference between the total energy 
and the total charge at the equilibrium configuration 
is always positive and of 
%. The magnitude of the difference is definitely 
$O(\epsilon^2)$ in our setup of the problem. 
Hence, if the effect of energy loss due to radiation is 
less than $O(\epsilon^2)$, there is no possibility of overcharging. 
To the contrary, if the energy loss is as large as 
$O(\epsilon^2)$, all parameter region in principle 
may give an example of overcharging. Hence, in the succeeding 
discussion we will not restrict ourselves to the case with $b=O(1)$. 
From the above correspondence, we understand that, 
when $b$ is of higher order, so is $g$, which corresponds to the 
case that $\ell$ is greater than $O(\epsilon)$. 
In this case, the difference in the rest mass and the charge 
of the point particle $a - c$ is also higher order and the proper 
separation between the horizon and the particle is as large 
as $O(|\log \epsilon|)$.

%%%%%%%%%%%%%%%%%%%%%%%%%%%%%%%%%%%%%%%%%%%%%%%%%%%%%%%%%%%%%%%
\subsection{Gauge invariant perturbation of the\ RN black hole excited 
by a charged particle}
\label{subsec:perturbation}
%%%%%%%%%%%%%%%%%%%%%%%%%%%%%%%%%%%%%%%%%%%%%%%%%%%%%%%%%%%%%%%
To estimate the energy emitted to infinity by the electromagnetic
and gravitational radiation, we use 
linear perturbation theory, which has been developed 
for a RN spacetime in Refs.~\cite{Kodama_04, Zerilli_75}.  
Here we adopt the formulation developed by
Kodama and Ishibashi~\cite{Kodama_04},  
in which the equations are reduced to
a set of decoupled second-order differential equations. 
%This property is very useful for our analysis. 

Owing to the background spherical symmetry, 
one can decompose the perturbations into scalar- and vector-types in general. 
It is easy to check that the
energy-momentum tensor of a charged particle in a radial orbit 
is purely of scalar-type, and therefore only the scalar-type perturbations are
excited. 
In addition to that, by choosing the coordinates so that 
the trajectory is along the axis we can restrict
our analysis to the axisymmetric mode ${\rmx m}=0$, where ${\rmx m}$ is 
the azimuthal index of spherical harmonics $Y_{\rmx l \rmx m}(\theta,\phi)$. 
As shown in Ref.~\cite{Kodama_04}, the
equations of the scalar-type perturbation are reduced to two decoupled
equations for two gauge invariant variables, $\Phi_\pm$.
They are given in terms of Fourier-harmonics expansion as
\begin{equation}\label{MasterEq}
 \frac{d^2 \Phi_{\pm} (r^{\ast})}{dr^{\ast 2}} + 
 (\omega^2 - V_{\pm}(r)) \Phi_{\pm} (r^{\ast}) 
= S_{\pm} (r^{\ast},\omega)~, 
\end{equation}
where the tortoise coordinate $r_*$ is defined by $dr_*/dr=1/f(r)$,
when $f(r)$ is as defined in Eq.~\eqref{RNf}, and $\omega$ is the frequency.
For simplicity, we abbreviate the labels 
$\rmx l, \rmx m,\omega$ to be attached to the master variables, 
effective potentials and source terms.
The effective potentials, $V_{\pm}(r)$, are defined by
\begin{eqnarray}\label{potential-scalar}
 V_{\pm}(r) := \frac{f(r)~ U_{\pm}(r)} {64 r^2 H^2_{\pm}(r)}~, 
\end{eqnarray}
with auxiliary functions 
\begin{eqnarray}
\displaystyle  H_{+}  := 1 - 3\delta x~, 
\quad  H_{-} := m + 3\rho x~,
\qquad  \delta:= \frac{1}{2 m} \left( \frac{\nu} {M} -1 \right)~,
\quad   \nu^2 := M^2 + \frac{4 m Q^2}{9}~,
\quad  \rho  := 1 + m \delta~,
\end{eqnarray}
where $x:= 2M / r$ and $z:= Q^2 / r^2$. 
Here we introduced $m := {\rmx l} ({\rmx l} + 1)-2$ 
where ${\rmx l} ({\rmx l} + 1)$ is the eigenvalue of the Laplace operator on
$S^2$.  
(It should be stressed that $m$ is different from the azimuthal
index ${\rmx m}$). 
The functions $U_{\pm}(r)$ in Eq.~\eqref{potential-scalar} are given by
\begin{eqnarray}
U_{+}(r) &\!\!:= &\!\!
- 2592 \delta^3 \rho x^4 + 576 ( 3 m \delta + 4 ) \delta^2 x^3
%\nonumber \\ && 
- 192 \delta x ( 3 x + m ) + 64 ( m + 2 )~, \nonumber \\
U_{-}(r) &\!\! := &\!\!
- 2592 \delta \rho^3 x^4 - 576 ( 3 m \delta -1 ) \rho^2 x^3
%\nonumber \\ && 
+ 192 m \rho x ( 3 x + m ) + 64 m^2 ( m + 2 )~.
\end{eqnarray}
The tensor composed of the scalar-type harmonics  
vanishes for ${\rmx l} = 1$, or equivalently for $m = 0$. 
This mode is called an exceptional mode in Ref.~\cite{Kodama_04}, 
which requires special treatment. In this case, 
although the master equation~\eqref{MasterEq} is 
still valid, the dynamical variable is $\Phi_{+}$ only. 
For the exceptional mode, $\delta$ takes the value 
$Q^2/9M^2$, which is obtained by taking the limit $m\to 0$. 

The source terms $S_{\pm} (r^{\ast},\omega)$  
are constructed from the energy-momentum tensor and the current induced
by a charged particle.
They are given by 
\begin{equation}
S_{\pm}(r^{\ast},\omega) := 
a_{\pm}(r)S_{\Phi} (r, \omega) + b_{\pm} S_{\mathcal{A}} (r,\omega), 
\end{equation} 
with the coefficients defined by
\begin{eqnarray}\label{ab-coeff}
  \{ a_+(r),~b_+ \} &\!\!:=&\!\! 
  \left\{ \frac{mQ}{2} + \frac{3 (M+\nu) Q}{2r},~
  \frac{3 (M+\nu) \kappa}{\sqrt{2}} \right\}~, \cr
  \{ a_-(r),~b_- \} &\!\!:=&\!\! 
  \left\{ 3(M + \nu) - \frac{4 Q^2}{r},~{-4 \sqrt{2}} Q \kappa \right\},
\end{eqnarray}
and
\begin{eqnarray}
\displaystyle  S_{\Phi} (r,~\omega) &\!\! :=&\!\!
\frac{\sqrt{2}f Q \kappa}{r^3 H} 
\left[ \left\{ -\frac{P_{S1}}{H} + 2( m + 2) \right\} 
\frac{\tilde{J}_t}{i\omega} \right]
\nonumber \\ &&
+ \frac{f}{r H}
\left( \frac{P_{S3}}{H} \frac{r S_t^r}{i \omega} 
            + 2 r^2 \frac{1}{i\omega}\frac{\partial S_t^r}{\partial r} 
            + 2 r^2 S_r^r \right)~,
\label{source_Phi} \\
S_{\mathcal{A}} (r,~\omega) &\!\! :=&\!\!
- \left( \frac{8 z f^2}{r^2 H}  
- \omega^2 \right) \frac{\tilde{J}_{t}}{i\omega}
   - f \frac{\partial}{\partial r}(f \tilde{J}_{r}) 
   -    \frac{ 2 \sqrt{2} f Q}{i \omega H \kappa} \frac{S_{t}^{r}}{r}~.
\label{source_A}
\end{eqnarray}
Here functions $H,~P_{S1}$ and $P_{S3}$ are defined by 
\begin{eqnarray}\label{source-aux}
  H       := m + 3 x -4 z~, \quad
  P_{S1}  := 4 x [ 2 z - 3 x + 6 + m ( m + 4 ) ]~,\quad
  P_{S3}  := 4   ( 3 x - 8 z )~,
\end{eqnarray}
$S_a^b(t)$ and $\tilde{J}_a(t)$ are respectively 
the Fourier components of the energy momentum tensor 
and the charge current of a radially falling charged particle 
that starts with a stationary point $r=r_0$ in the infinite past 
$t \to -\infty$. 
Their explicit forms are expressed as 
\begin{eqnarray}
\label{source-table}
%----------------------------------------------------------------------
%%%% S_t^{r} and S_r^{r} %%%%%%%
S_t^{r} &=& 
- \kappa^2  \mu f \left( \frac{dT}{ds} \right) 
\frac{ e^{i \omega T} }{r^2}  Y_{ {\rmx l} 0}\,\theta(r_0-r), \qquad
S_r^{r} = 
 \kappa^2  \mu f^{-1} \left( \frac{dR}{ds} \right)
 \frac{e^{i \omega T}}{r^2}  Y_{ {\rmx l} 0 }\,\theta(r_0-r), \cr
%%%%% \tilde{J} %%%%%%
\tilde{J}_t &=&
\frac{q}{ {\rmx l} ({\rmx l} + 1) } e^{i \omega T} Y_{{\rmx l} 0}\,\theta(r_0-r), \qquad
\tilde{J}_r =  
- \frac{q}{ {\rmx l} ({\rmx l} + 1)  } \left( \frac{dT}{dR} \right) 
e^{i \omega T} Y_{{\rmx l} 0}\,\theta(r_0-r),
%----------------------------------------------------------------------
\end{eqnarray}
where $Y_{{\rmx l} 0} = \sqrt{(2 {\rmx l} + 1) / 4 \pi}$.
%Here the step function appears 
%because the trajectory of the particle is restricted to 
%the range $r < r_0$. }
In these expressions $dT/ds$, $dR/ds$, 
$dT/dR:=(dT/ds)/(dR/ds)$ and $T$ are to be understood as functions 
of $R$, and $R$ is replaced with $r$. 
The function $T$ is obtained by 
integrating Eqs.~(\ref{dR-ds}) and (\ref{dT-ds}) and eliminating $s$.

The formal solutions $\Phi_{\pm}$ of Eq.~\eqref{MasterEq} 
applicable to both the generic and exceptional modes 
are obtained by the usual Green function method.
The asymptotic form of the solution at $r \to +\infty$ is given by 
\begin{eqnarray}
\Phi_{\pm}(r^{\ast}, \omega) =  
\frac
{e^{i \omega r^{\ast}} {\mathcal{X}}_{\pm}(\omega)}
{W[\Phi_{\pm}^{\rm up}, \Phi^{\rm{in}}_\pm]}~,
%\quad \textrm{for}~ r \to +\infty, 
\end{eqnarray}
with 
\begin{eqnarray}
{\mathcal{X}}_{\pm}(\omega) :=
\int^{r_0^{\ast}}_{r_+^{\ast}}
\Phi_{\pm}^{\rm{in}} (r^{\ast}, \omega) 
S_{\pm}(r^{\ast}, \omega) d r^{\ast},
\label{inhomo-sol}
\end{eqnarray}
where the functions $\Phi_{\pm}^{\rm{up/in}} (r^{\ast}, \omega)$
are the homogeneous solutions of Eq.~\eqref{MasterEq} that satisfy
the boundary conditions
$\Phi_{\pm}^{\rm{up}} (r^*,~\omega) \to  e^{ i \omega r^*}$
at $r^\ast\to\infty$ 
and
$\Phi_{\pm}^{\rm{in}} (r^*,~\omega) \to  e^{- i \omega r^*}$
at $r^{\ast} \to -\infty$, respectively. 
We also introduced the Wronskian defined by 
$W[\Phi^1, \Phi^2] := (\partial_{r^*} \Phi^1(r^*)) \Phi^2(r^*)
- \Phi^1(r^*) (\partial_{r^*} \Phi^2(r^*))$.

The energy flux carried by electromagnetic and gravitational waves
to infinity is described in terms of the variables $\Phi_{\pm}$ 
as given in Eq.~(\ref{Flux-boundary-G}) 
in Appendix~\ref{sec:flux}. 
In this expression 
$\bar \Phi_{\pm}(r,t)$ are the complex conjugations of
$\Phi_{\pm}(r,t)$. 
Substituting the solution~\eqref{inhomo-sol}
into the time integral of Eq.~\eqref{Flux-boundary-G}, 
with the aid of Parseval's theorem, 
we obtain the total energy radiated to infinity
\begin{eqnarray}
E_{ \infty } &\!\! =&\!\! 
\int_{0}^{+\infty} d\omega \sum_{\rmx l} 
\frac{ 8 \pi { {\rmx l}({\rmx l}+1) } \omega^2}
{9 \kappa^2 \nu ( M + \nu )}
  \left({|{\cal X}_{+}|^2\over |W_+|^2} + {({\rmx l}-1)({\rmx l}+2) \over 16} 
{|{\cal X}_{-}|^2\over |W_-|^2}
\right)~,
\label{Einfty}
\end{eqnarray}
where $W_\pm:={W[\Phi_{\pm}^{\rm up}, \Phi^{\rm{in}}_\pm]}$.
The explicit form of ${\cal X}_{\pm}$ is given in Eq.~(\ref{Integrands}). 
%In Appendix~\ref{sec:flux} it is also explained that the last two terms in 
%Eq.~(\ref{Integrands}) can be rewritten as Eq.~(\ref{lasttwoterms}). 
We point out that the integrals in the expressions for  ${\cal X}_\pm$ 
given in Eq.~(\ref{Integrands}), 
with the last two terms replaced with Eq.~(\ref{lasttwoterms}) 
all take the form 
\begin{equation}\label{schematic-Integral}
\int_{r_+}^{r_0} I(r)\Phi^{\rm in}_\pm(r) e^{i\omega T(r)} dr, 
\end{equation}
where $I(r)$ is a certain regular function of $r$, whose typical 
scale of variation $\Delta r$ is $O(M^{-1})$.
For large $\omega$, the integrals~\eqref{schematic-Integral} 
take the structure of rapidly oscillating function 
$\Phi^{\rm in}_\pm(r) e^{i\omega T(r)}$ multiplied by the 
slowly varying function $I(r)$. In general, such an integral is known 
to be exponentially suppressed for large $M \omega$. Thus, the 
$\omega$-integral in Eq.~\eqref{Einfty} has an effective high 
frequency cutoff at $O(M^{-1})$. 
%%%%%%%%%%%%%%%%%%%%%%%%%%%%%%%%%%%%%%%%%%%%%%%%%%%%%%%%%%%%%%%%%%%%

Now we show that this radiated energy is 
$O(\epsilon^4)$ or higher. It is obvious that $|{\cal X}_\pm(\omega)|$ 
is suppressed by a factor of $O(\epsilon)$ since each term 
manifestly contains $\mu$ or $q$. 
Therefore what we have to show is that 
there is an additional suppression of $O(\epsilon)$ in 
$|{\cal X}_\pm(\omega)|$.

As a preparation, we show that the inverse of the Wronskian 
is suppressed in the limit $\omega\to 0$ like 
$W_\pm^{-1} \propto (M\omega)^{\rmx l}$. 
The presence of this suppression is understood as follows. 
Since the Wronskian is constant independently of $r$, 
we evaluate it in the limit $r \to \infty$ as
\begin{equation}
\label{Wronskian}
W_\pm =\lim_{r\to\infty} 
\left( 
i \omega \Phi_{\pm}^{\rm{in}}(r) -
\frac{\partial \Phi_{\pm}^{\rm{in}}(r)}
{\partial r}  
\right) 
~e^{i \omega r}~.
\end{equation}
Note that in the limit $r \to \infty$, the difference between $r^{\ast}$ and 
$r$ is suppressed by $1 / r$, so that we can use $r$ instead of $r^{\ast}$ 
in Eq.~\eqref{Wronskian}.
When $r$ is large enough compared with $M$, the homogeneous equation 
corresponding to Eq.~\eqref{MasterEq} can be approximated as
\begin{equation}
\label{flat-Meq}
\frac{d^2 \Phi_{\pm} (r)}{dr^{2}} + 
 \left( \omega^2 - \frac{\rmx l(\rmx l+1)}{r^2} \right) \Phi_{\pm} (r) = 0~.
\end{equation}
The general solution of this equation, 
which is written in terms of the Bessel function of the first kind 
$J_{\rmx l}$ and that of the second kind $Y_{\rmx l}$,
should also describe $\Phi^{\rm in}_{\pm}$.
Hence, we have 
\begin{equation}
\Phi^{\rm in}_{\pm}
\approx C_J \sqrt{\pi \omega r\over 2} J_{{\rmx l} + 1/2} (r
 \omega) +
C_Y \sqrt{\pi \omega r\over 2} Y_{{\rmx l} + 1/2} (r \omega)~,
\label{Bessel-exp}
\end{equation}
where $C_J$ and $C_Y$ are coefficients to be determined by the 
condition imposed near the horizon. 
Then, for a small $r$ of $O(M)$ on the verge of the validity of this 
approximate solution, the two terms in Eq.~(\ref{Bessel-exp}) should be 
equally important in general. Furthermore, as there is no significant 
feature in the potential, the amplitude of these terms must be $O(1)$. 
This determines the order of 
magnitude of the coefficients as $C_J=O((M \omega)^{-{\rmx l}-1})$ 
and $C_Y = O((M \omega)^{{\rmx l}})$, since 
for a small $r \omega$
the above expression asymptotically behaves as 
\begin{equation}
\Phi^{\rm in}_{\pm}
\to C_J 
\sqrt{\pi \over 2}\left({\omega r\over 2}
\right)^{{\rmx l}+1} {1\over \Gamma\left({\rmx l}+{3/2}\right)}
- C_Y 
{1 \over \sqrt{2\pi} }\left({\omega r\over 2}
\right)^{-{\rmx l}} \Gamma\left(-{\rmx l}-{1/2}\right)~. 
\end{equation}
On the other hand, for a large $r$, the first term dominates to give 
\begin{equation}
\Phi^{\rm in}_{\pm}
\to C_J \cos\left(\omega r-{\pi\over 2}({\rmx l}+1)\right). 
\end{equation}
Substituting this expression into Eq.~(\ref{Wronskian}), 
the order of magnitude of the Wronskian is estimated as 
$W_\pm=O\left({1\over M}(M\omega)^{-\rmx l}\right)$. 
As anticipated, we find that the inverse of the Wronskian scales 
like $\propto (M\omega)^{\rmx l}$.

Now, in order to prove the presence of an additional suppression factor 
in $|{\cal X}_\pm(\omega)|$, 
we focus on the fact that 
$(dR/ds)^2=V^{(\rm{o})}$ 
is always suppressed in the present setup. 
It will be obvious that 
$V^{(\rm{o})}(r)$ 
is a quadratic function of $1/r$ bounded from below. 
Furthermore, we know that both $V^{(\rm{o})}$ and 
$dV^{(\rm{o})}/dr$ vanishes at $r=r_0$. 
Therefore $V^{(\rm{o})}$ takes its maximum value 
 at $r=r_+$, in the interval of our interest between $r_+$ and $r_0$.
Hence, we have  
\begin{equation}
 V^{(\rm{o})}(r)\leq V^{(\rm{o})}(r_+) = 
 {1\over \mu^2} \left( E - { q Q \over M + \sqrt{M^2 - Q^2}} \right)^2
 = {4(a - c)^2 \over c^2}\epsilon^2 + O(\epsilon^3).  
\end{equation}
This implies that the velocity of a particle 
$dR/ds$ given in Eq.~(\ref{dR-ds}) is always at most $O(\epsilon)$. 
%(When $b$ and hence $a-c$ are small, 
%$dR/ds$ is further suppressed.) 
When the particle moves very slowly, the amount of emitted 
radiation is also expected to be small. 
This intuition can be made explicit in 
the expression for ${\cal X}_\pm$ as follows. 
One can see that each expression in Eq.~\eqref{schematic-Integral} 
is regular on the boundaries of the integral. 
Here we replace $e^{i\omega T(r)}$ with an equivalent expression
\begin{equation}
\label{subs}
{f \over i\omega}\left({dR\over ds}\right)\left(f{dT\over ds}\right)^{-1} 
\left(  \frac{ \partial e^{i\omega T(r)} }{\partial r} \right),  
\end{equation}
and perform integration by parts. 
The regularity on the boundaries of the integral is not ruined 
thanks to the presence of a factor $f$ and 
$dR/ds$ in the above expression (\ref{subs}). 
(Notice that the combination $f(dT/ds)$ is regular on the horizon.)
This manipulation adds at least 
one $dR/ds$ factor at the expense of decreasing the power of $\omega$ by
one. Repeated application of this integration by parts is restricted 
by the requirement for the convergence of the $\omega$-integral 
in Eq.(\ref{Einfty}). 
However, this is not such a severe constraint, because the inverse
Wronskian squared $|W_\pm|^{-2}$
gives suppression for a small $\omega$ proportional to $\omega^{2{\rmx
l}}$ as mentioned above. 
Therefore this additional suppression owing to the $dR/ds$ factor guarantees 
that the total energy emitted 
to infinity is at most $O(\epsilon^4)$. 

To summarize, the total energy emitted 
to infinity by a particle that falls from the unstable stationary point to 
the horizon,
$E_{\infty}$, is always suppressed, and it cannot 
be as large as $O(\epsilon^2)$ for any parameter choice. 
Hence, the effect of the energy loss due to radiation 
does not affect the inequality~(\ref{eq:difference-EC}), 
which is a relation at the level of $O(\epsilon^2)$. 
Thus, we conclude that 
\begin{equation}
\label{Final-result}
E_{\rm eq} - E_{\infty} > q + Q, 
\end{equation}
is always satisfied. Namely, 
the total mass of the final state can never be reduced 
below the critical value that corresponds to the extremal bound.  

%%%%%%%%%%%%%%%%%%%%%%%%%%%%%%%%%%%%%%%%%%%%%%%%%%%%%%%%%%%%%%%   Sec 
%\section{Conclusion}
%%%%%%%%%%%%%%%%%%%%%%%%%%%%%%%%%%%%%%%%%%%%%%%%%%%%%%%%%%%%%%%   Sec 
\section{Conclusion}
\label{sec:conclusion}
In this work, we have examined the back reaction effects of
$O(\epsilon^2)$ when a charged particle whose mass and charge are 
of $O(\epsilon)$ is absorbed by a nearly extremal RN black hole. 
To avoid the technical difficulties related to the electromagnetic and 
gravitational self-force, 
we concentrated on the case of the marginal orbit, 
which is the separatrix dividing the plunge and recoil orbits. 
%In contrast to the previous results in the test particle approximation, 
We first showed that, with the aid of 
an exact solution, the total 
energy of the system is always greater than the total charge for 
the equilibrium configuration that the marginal orbit passes through.  
Then, we demonstrated that the 
radiative energy loss as the particle is falling into the black hole 
from the equilibrium position is $O(\epsilon^4)$ or higher. 
Combining these results, we succeeded in proving that 
the total energy of the system composed of 
a charged particle and a black hole 
is always greater than their total charge for the marginal orbit, 
once back reaction effects are properly taken into account.  
In short, the back reaction effects 
prevent a nearly extremal RN black hole from being overcharged, and hence 
we conclude the cosmic censorship conjecture is not violated. 
%According to our analysis, there is no strong supporting evidence to 
%violate the cosmic censorship in Hubeny's overcharging process.
%Before closing our paper, however, we firstly comment on the validity of 
%the key assumptions used in our analysis. 

As was mentioned in Sec.~\ref{subsec:Backreaction}, 
our discussion relies on the assumption that the marginal orbit 
passes through the unstable stationary configuration. 
Although it is difficult to imagine that this is not the case, 
our modest conclusion at the moment will be that there is no 
evidence that supports the possibility of overcharging a nearly 
extremal RN black hole by absorbing a charged particle.  
If we further accept a rather natural assumption that the final state 
of all the plunging orbits has a higher energy than the case of 
the marginal orbit for given any charges of particle and black hole, 
we can exclude the possibility of overcharging without the restriction to 
the marginal orbit.

To obtain a definitive answer to the question whether these  
assumptions are correct or not, it is necessary to solve 
the orbital evolution directly including the self-force effects. 
The formal expressions for the gravitational self-force in vacuum spacetime 
was already derived by Mino, Sasaki and Tanaka, 
and independently by Quinn and Wald~\cite{MiSaTaQuWa},   
%the resulting particle's modified equation of motion 
and is known as the MiSaTaQuWa force. 
%Removing the restriction that the MiSaTaQuWa equation is only valid
%in the vacuum space time, and make it applicable to the general 
%curved space time with external fields, we need to generalise the MiSaTaQuWa 
%equation to the electro-vacuum case. 
%However, the actual evaluation of the self-force in the Kerr
%background, especially when we consider the application to the overcharging pro%blem, 
%still requires further investigation. 
In the case of a charged particle in the non-vanishing electro-magnetic
background, even the fundamental formulation corresponding to the
MiSaTaQuWa force is lacking. 
Although there is a recent work in this direction~\cite{Futamase:2008zz}, 
it has not yet been developed to the level applicable to the overcharging 
problem discussed in this paper. 

The main lesson of this paper is 
that it is important to include the non-dissipative part of the self-force 
when we examine the possibility of overcharging. 
In the context of the spinning up of a Kerr black hole proposed in 
Ref.~\cite{Jacobson:2009kt}, 
Barausse et al. have shown that, 
to avoid overspinning, it is not sufficient to take into account 
only the dissipative part of the self-force, \ie \ 
the energy loss and the angular momentum loss due to gravitational
radiation~\cite{Barausse_10}.
As was discussed in Ref.~\cite{Barausse_10}, also in this case 
it is essential to take into account the non-dissipative part 
of the self-force. 
However, in contrast to the overcharging process discussed in this paper, 
we cannot expect the existence of the stationary intermediate configuration 
characterizing the marginal orbit. 
Therefore, directly analysis of the self-force will be unavoidable.

%%%%%%%%%%%%%%%%%%%%%%%%%%%%%%%%%%%%%%%%%%%%%%%%%%%%%%%%%%%%%%%%%%%%%
%%%%%%%%%%%%=============================================%%%%%%%%%%%%

\acknowledgments
It is our pleasure to thank Tetsuya Shiromizu for his valuable comments,
especially for pointing out 
the potential existence of subtle issues in using the DRN solution 
as we discussed in Appendix~\ref{sec:singularity}. 
We also wish to thank Akihiro Ishibashi, Masashi Kimura
Shunichiro Kinoshita, Norihiro Tanahashi and Chul-Moon Yoo 
for fruitful discussions.
We also grateful to Jonathan White for his careful reading of 
the manuscript, which is very useful to improve the presentation.
%help with English corrections of our paper}.
N.~S. and T.~T. acknowledge support by the Grant-in-Aid for Scientific
Research (No.\ 21244033).
This work was supported by the Grant-in-Aid for the Global COE Program
"The Next Generation of Physics, Spun from Universality and Emergence"
from the Ministry of Education, Culture, Sports, Science and Technology
of Japan.

%%%%%%%%%%%%%%%%%%%%%%%%%%%%%%%%%%%%%%%%%%%
%%                APPENDIX           %%%%%%
%%%=========================================
\appendix
%%%====================================
%%%%%%%%%%%%%%%%%%%%%%%%%%%%%%%%%%%%%%%%%%%

%------------------------------------------------------------------------------

%%%%%%%%%%%%%%%%%%%%%%%%%%%%%%%%%%%%%%%%%%%%%%%%%%%%%%%%%%%%%%%%%%%%%
%%===========================================================
\section{Geometry near the singularity in the DRN solution}
\label{sec:singularity}
%%%%%%%%%%%%%%%%%%%%%%%%%%%%%%%%%%%%%%%%%%%%%%%%%%%%%%%%%%%%%%%%%%%%%

In Sec.~\ref{subsec:Remark-DRN},
we implicitly assumed that the singularity in the DRN solution 
is the counterpart of a charged particle. 
The aim of this appendix is to confirm this correspondence. 
The main focus is on the asymptotic behavior near the singularity,  
so as to show that the deformation from the 
spherically symmetric singularity is sufficiently small. 

To analyze the geometry near the singularity in the DRN solution, 
the original $(\rho,z)$ coordinates are inappropriate since 
they do not cover the region where the singularity resides.  
In order to analytically continue DRN
metric~(\ref{double_RN_metric}),
we can use the bipolar coordinates $(r_1, \theta_1)$
defined by Eqs.~\eqref{bipolar}. 
However, to see the behavior near the singularity, 
it is more convenient to change the radial coordinate from 
$r_1$ to $\cal{D}$. Furthermore, 
as we are focusing on the behavior near the 
singularity, we use the rescaled variable $\Delta$ 
defined by ${\cal{D}} = \Delta \epsilon^2$, instead of ${\cal D}$. 
For simplicity, we concentrate on the case that the stationary point is 
rather close to the black hole horizon, in which $g$ is $O(1)$. 
In the case with a smaller $g$, which corresponds to a larger $\ell$, 
the deformation is expected to be even weaker. 

When $g$ is $O(1)$, it is more convenient to use the parametrization 
$\lambda=\tilde g/g$ instead of $\tilde g$ introduced in
Eqs.~(\ref{e-parameter2}). 
Hence, our parametrization adopted here is  
\begin{equation}
 e_1    =  a \epsilon~,\quad
 e_2    =  1~,\quad
 \gamma =  g \epsilon^2,\quad
 \ell   =  \lambda \epsilon.
\end{equation} 
The other parameters $m_1, m_2, \sigma_1^2$ and $\sigma_2^2$ are 
to be determined from Eqs.~(\ref{def-sigma-double-RN}), 
\eqref{def-gamma-double-RN} and \eqref{balance-eq} to the sufficiently 
high order in $\epsilon$. For the present purpose, 
the approximate expressions~\eqref{expanded-variables-DRN} are not 
sufficient.

To write down the metric and the electric potential 
in the $(\Delta, \theta_1)$ coordinates, we need to describe 
$r_1$, $r_2$ and $\theta_2$ in terms of $(\Delta, \theta_1)$. 
From the definitions~\eqref{bipolar} and \eqref{double_RN_D}, 
setting $z_1 = 0$ and $z_2 = \ell$, 
we obtain 
\begin{eqnarray}
\label{R2vsTheta2}
r_2                    &\!\! =&\!\! 
m_2 + \frac{ (r_1 - m_1) {\rm{cos}}{\theta}_1 - \ell }{\cos\theta_2}~,\cr
\sin^2 \theta_2  &\!\! =&\!\! 
\frac{ [ (r_1 - m_1)^2 - \sigma_1^2 ] ~{\rm{sin}}^2 \theta_1 }
{(r_2 - m_2)^2 - \sigma_2^2}~, \cr
r_1 &\!\! = &\!\! r_2^{-1}\left[
(e_1 - \gamma -\gamma \cos\theta_2)(e_2+\gamma-\gamma \cos\theta_1)+{\cal D}
\right]~, 
\end{eqnarray}
which can be solved by iteration in the presented order, 
once we have appropriate initial values for $r_1$ and $\theta_2$. 
The location of the singularity is specified by ${\cal D}=0$, 
where the lapse function $g_{tt} = - H$ and the electric potential $\Phi$
diverge.  This point  
is expected to be close to the critical spheroid, and hence we have
$r_1\approx m_1$.  
Then, we also have $\theta_2 \approx -\pi$ near the singularity. 
Using these crude estimates as the initial condition for the iteration, 
we can solve these equations to a sufficiently high order in $\epsilon$.

Since the coordinate $\Delta$ is not 
best suited for seeing how the DRN solution deviates from 
the single RN singularity, it is better to further replace
the radial coordinate to a physically well-motivated one. 
One simple possibility is to adopt the inverse of the electric potential 
as the radial coordinate, \ie
\begin{equation} 
\label{physical-R}
 R := {a \tilde c \epsilon^2\over (\Phi-\Phi_0)}~, 
\end{equation}
where $a\epsilon$ is the charge of the particle 
and we will find that the factor $\tilde c\epsilon$ 
takes care of the change of the time coordinate,
\begin{equation}
 dT := \tilde c \epsilon  dt, 
\end{equation}
with $\tilde c$ being a constant of $O(1)$ to be determined later, 
once we notice that the combination $A_t dt (= - \Phi dt)$ is 
gauge invariant. 
Here, the new time coordinate $T$ appropriate to describe the charged
particle as a perturbed RN geometry elapses slower than the time of 
an asymptotic observer $t$ by the red shift factor $\tilde c\epsilon$ since 
the particle is staying deep inside the gravitational potential of the black hole.
Also, a constant 
$\Phi_0$ was introduced to adjust the zero point of the electric potential.
$\Phi_0$ will be also determined later. 
To focus on the range where $e_1/R=a\epsilon/R=O(1)$, 
we mainly use the rescaled radial coordinate $\varrho=R/\epsilon$, 
assuming $\varrho = O(1)$.
(Notice that the coordinate $\varrho$ is different
from $\rho$ in the cylindrical Weyl coordinates, used to describe the
DRN solution in Eqs.~\eqref{double_RN_metric} and \eqref{double_RN_EM}.)

Then, it is straightforward to expand
the metric in these coordinates up to $O(\epsilon)$. 
By choosing 
\begin{eqnarray}
\tilde c &\!\! = &\!\! {\sqrt{a\lambda(a\lambda -g)}\over a}
  - \epsilon {\lambda(g^2-2ag\lambda+2a^2\lambda^2)\over
    2a \sqrt{a\lambda(a\lambda -g)} }+O(\epsilon^2),~\cr
\Phi_0 &\!\! = &\!\! 
1 - \lambda \epsilon 
+ \epsilon^2 {\lambda (2a \lambda - g ) \over 2a} + O(\epsilon^3)~,
\end{eqnarray}
the resulting metric is expressed in the following form
\begin{eqnarray}
\label{R1vsD}
g_{TT}&\!\! =&\!\! - \tilde f(\varrho) 
\left[ 1 - \epsilon \cos\theta_1 \left( 
\frac{2 g \varrho}{ \sqrt{a \lambda (a \lambda - g)}} \right)    
+ O(\epsilon^2) \right], \cr
%%----------------------------------------------------------
g_{RR}&\!\! =&\!\! {1\over \tilde f(\varrho)} \left[
1 + \epsilon \cos\theta_1 
 \left\{ \frac{6(a^2 + \varrho^2)} {a} - { 2 \varrho (6a \lambda - 5g) \over
  \sqrt{a\lambda(a\lambda -g)}} \right\}
+ O(\epsilon^2) \right], \cr
%-------------------------------------------------------------
g_{ij} &\!\! = &\!\! 
 R^2 \gamma_{ij}\left[1 + 2 \epsilon \cos\theta_1 
 \left\{ {\lambda(a^2 + \varrho^2)
   - 2 \varrho \sqrt{ a \lambda (a \lambda - g) }\over a \lambda} \right\}
   + O(\epsilon^2)\right], \cr
g_{R\theta_1} &\!\! = &\!\! \epsilon R^2 \sin\theta_1 \left(
  {\varrho\over a} - \sqrt{a\lambda\over (a \lambda - g )}\right)
 +O(\epsilon^2)~,
\end{eqnarray}
where we have defined 
\begin{eqnarray}
\label{final-f}
\tilde f &\!\! = &\!\! {a^2\over \varrho^2}
-\left(
   {2\sqrt{a\lambda(a\lambda -g)}\over \lambda}
     - \epsilon {g(2a\lambda-g)\over
       \sqrt{a\lambda(a\lambda -g)}}  \right){1\over \varrho} + 1
        = 
        \frac{q^2}{R^2} - \frac{ 2 \mu }{R} + 1~.
\end{eqnarray}
The second equality gives the expressions of $q$ and $\mu$ in terms of 
$a$, $g$, $\lambda$ and $\epsilon$, which are consistent with 
$q = a \epsilon,~\mu = c \epsilon$ in 
Eqs.~(\ref{e-parametrization-particle+RN}) supplemented with the relations
 $\lambda g \equiv \tilde g=4a+O(\epsilon)$ 
and  $g  =  2 \sqrt{a^2 - c^2}+O(\epsilon)$ in Eqs.~(\ref{mapping}).
In the next leading order in $\epsilon$ the metric given in 
Eqs.~(\ref{R1vsD}) has only dipole type perturbation 
at around $R=O(\epsilon)$. Perturbations belonging to higher multipoles 
are of $O(\epsilon^2)$ or higher.

The asymptotic behavior of perturbations for small $R$ 
is to be determined in line with the nature of a charged 
particle. For example, when an external 
dipole field is imposed, a charged particle will gain 
some induced dipole moment. However, the magnitude 
of the induced dipole moment will depend on the property of the 
particle to some extent. 
In this sense there remains arbitrariness in 
the most relevant choice of the 
boundary conditions for our charged particle. 
Instead of discussing this, we show 
that the effect of the induced dipole of $O(\epsilon)$ in 
the expressions~(\ref{R1vsD})  
does not contribute to 
the total energy of the system at the level of our interest,  
$O(\epsilon^2)$.

The key observation is that it is 
at $R=O(\epsilon)$ where the dipole perturbation is $O(\epsilon)$.
This dipole field is a superposition of the external 
field caused by the presence of a charged black hole and 
the induced dipole of the point particle. Hence, the latter amplitude 
will be also at most $O(\epsilon)$. 
The dimensionless perturbation 
caused by an object having a dipole moment $D$ is $O(D/R^2)$. 
Therefore the magnitude of the dipole moment is evaluated 
as $D=O(\epsilon^3)$. In the region $R\gg O(\epsilon)$, 
the configuration can be understood as a RN black hole solution perturbed 
by a charged particle and an electric dipole placed 
at the same place. In such a regime the effect due to the 
dipole is at most $O(\epsilon^3)$. 
On the other hand, in the vicinity of $R=0$ we can evaluate 
the energy due to the dipole from the expanded metric presented 
in Eqs.~(\ref{R1vsD}). The explicit expanded metric shows that 
the monopole part of the perturbation starts with $O(\epsilon^2)$. 
Since the change of the mass $\delta \mu$ will appear in the 
monopole part of the dimensionless metric perturbation 
as $\delta\mu/R$, we find that $\delta\mu$ is at most 
$O(\epsilon^3)$ in the region $R\alt \epsilon$. 
As a result, we conclude that the contribution of the dipole 
to the total energy is $O(\epsilon^3)$. The higher 
multipoles can be discussed in a similar way, and we find 
that they are even higher order in $\epsilon$.

%%%%%%%%%%%%%%%%%%%%%%%%%%%%%%%%%%%%%%%%%%%%%%%%%%%%%%%%%%%%%%%%%%%
\section{$\epsilon$-expansion of the parameters in the DRN solution}
%~\eqref{e-parameter2}}
\label{sec:e-prameter}
%%%%%%%%%%%%%%%%%%%%%%%%%%%%%%%%%%%%%%%%%%%%%%%%%%%%%%%%%%%%%%%%%%%

This appendix is devoted to deriving the 
appropriate order of $\gamma$ and $\ell$ in terms of 
$\epsilon$-parametrization.
Our starting point is the assignment of the ordering 
that defines the setup of our problem given in Eqs.~(\ref{trial}).  
The parameters are also constrained by the inequalities 
$\sigma_1^2 < 0$ and $\sigma_2^2 > 0$, which are required because 
$\sigma^2$ is positive for a BH and negative for a singularity. 

From the ordering given in Eqs.~\eqref{trial}, 
the leading term of $\gamma$ in Eq.~\eqref{def-gamma-double-RN} 
can be read as
\begin{equation}
\gamma = \frac{e_1 - m_1}{1 + \ell} \left( 1 + O(\epsilon) \right)~.
\label{eq:order-gamma1}
\end{equation}
Substituting the above approximate $\gamma$ into the 
balance condition~\eqref{balance-eq}, we find 
\begin{equation}
\ell (e_1 - m_1) = O(\epsilon^2).
\label{eq:order-lem1}
\end{equation}
Multiplying $\ell$ on both sides of Eq.~\eqref{eq:order-gamma1},~ 
with the aid of~\eqref{eq:order-lem1}
we have $\ell \gamma=O(\epsilon^2)$. 
Next, using the last equality in Eqs.~\eqref{trial}, we find 
\begin{equation}
 e_1 - m_1 = \varpi + O(\epsilon^2). 
\label{e1m1delta}
\end{equation}
Plugging Eq.~\eqref{e1m1delta} into Eq.~\eqref{eq:order-lem1},
we also have $\ell \varpi = O(\epsilon^2)$. 

%We first consider the case with $\ell=O(\epsilon)$. 
%~\gamma = O(\epsilon)$ since $\gamma = O(1)$ 
%contradicts $\gamma = (e_1 - m_1)/ (\ell + m_2) = O(\epsilon)$.
We now show that Eq.~\eqref{eq:order-lem1} is in fact further suppressed.
Substituting Eq.~(\ref{def-gamma-double-RN}) 
into Eq.~(\ref{balance-eq}) and expanding it up to $O(\epsilon^2)$ 
after multiplication of the factor $(\ell+m_1+m_2)$ on both sides, 
and using Eq.~\eqref{eq:order-lem1} and the relation 
$\ell \varpi = O(\epsilon^2)$, we obtain
$2m_1(e_1-m_1-\varpi)+\ell(e_1 - m_1) = O(\epsilon^3)$.
With Eq.(\ref{e1m1delta}), this reduces to
\begin{equation}
\ell(e_1-m_1)=O(\epsilon^3),  
\end{equation}
which also implies $\ell \gamma = O(\epsilon^3)$ 
in the same manner as above and 
$\ell\varpi = O(\max(\epsilon^3, \ell \epsilon^2))$ with 
the aid of Eq.~\eqref{e1m1delta}. 
From Eq.~(\ref{def-gamma-double-RN}),  
\begin{equation}
 m_1 = \frac{ (1 + \varpi)(e_1 - \gamma) - \ell\gamma  } {1 + \gamma}, 
\end{equation}
follows. 
Substituting this expression for $m_1$ into 
$\sigma_1^2$ given in Eqs.~(\ref{def-sigma-double-RN}),  
with the aid of the relations~(\ref{eq:order-gamma1}) and (\ref{e1m1delta}),
we obtain
\begin{equation}
\sigma_1^2 =  \gamma^2 + O(\epsilon^4).
\end{equation}
Thus, the requirement that $\sigma_1^2<0$ can be consistent only when 
$\gamma=O(\epsilon^2)$. Then, immediately both $e_1-m_1$ and $\varpi$ 
turn out to be $O(\epsilon^2)$

To summarize, 
if $\ell$ is $O(\epsilon)$, we find that both 
$e_1 - m_1 = \gamma + O(\epsilon^3)$ and
$\varpi$ are $O(\epsilon^2)$. When $\ell$ is not as small as $O(\epsilon)$, 
the order of $e_1-m_1$ and $\gamma$ depends on the magnitude of $\ell$
in such a way that 
$\ell(e_1-m_1) = \ell (\ell + 1) \gamma + O(\epsilon^4)$ 
becomes $O(\epsilon^3)$. By contrast, $\varpi$ stays 
$O(\epsilon^2)$ irrespective of the magnitude of $\ell$.

%%%%%%%%%%%%%%%%%%%%%%%%%%%%%%%%%%%%%%%%%%%%%%%%%%%%%%%%%%%%%%%
\section{Mapping an exact static solution to 
 an equilibrium configuration of a charged particle in the RN spacetime}
\label{subsec:Mapping}
%%%%%%%%%%%%%%%%%%%%%%%%%%%%%%%%%%%%%%%%%%%%%%%%%%%%%%%%%%%%%%%
In this appendix we consider the mapping of the 
DRN static solution to an equilibrium configuration of
a charged particle in a RN black hole background: the BHP picture.
It is not so trivial to find the relations between 
the different descriptions, especially the relations 
between $\tilde g$ and $g$ in Eqs.~\eqref{mapping}, 
and the parameters 
$a$, $b$ and $c$ given in Eqs.~(\ref{e-parametrization-particle+RN}).  
To obtain these relations, we focus on two geometrical quantities:
one is the proper distance $L$ along the symmetric axis between the 
event horizon of the black hole and the equilibrium position of
the particle, and the other is the proper distance $D$ from the
event horizon to the asymptotic region, measured along the symmetric 
axis in the direction opposite to the particle.
For our current purpose, in linear perturbation theory analysis, 
it is enough to
estimate the emitted energy at the leading order in $\epsilon$ 
since the energy flux~\eqref{Einfty} is always $O(\epsilon^2)$. 
The back reaction effects on the orbit contribute only to the
higher order corrections in the energy flux.
Therefore we consider the mapping at the level of linear perturbation 
of a RN black hole, restricted to the order in $\epsilon$ expansion 
necessary to specify the setup in the BHP picture.

%--------------------------------------------------------------------------
We first compute the above geometrical quantities in the BHP picture. 
Since we consider the equilibrium state, however, the parameters 
$a,~b$ and $c$ in Eqs.~\eqref{e-parametrization-particle+RN}
are not independent but satisfy the relation
\begin{equation}
b = \sqrt{a^2 - c^2} + O(\epsilon),
\label{discriminant}
\end{equation}
which is derived from the equilibrium conditions $dR/ds=dR^2/ds^2=0$.
The value of the radial coordinate at the equilibrium position is also
derived from these equilibrium conditions as
\begin{equation}
r_{0} =  1 +
\frac{2a}{\sqrt{a^2 - c^2}}~\epsilon + O(\epsilon^2).
\label{charged-particle-distance}
\end{equation}
Then, the quantities mentioned above are calculated in the BHP picture as 
\begin{eqnarray}
L &\!\! =&\!\! \int_{r_+}^{r_0} \sqrt{g_{rr}} dr'
 = \int_{r_+}^{r_0} \frac{dr'}{\sqrt{f(r')}}
 = \frac{1}{2} \log \frac { a + c } {a - c} + O(\epsilon), 
\label{dL-perturbative} \\
D(r) &\!\! =&\!\! \int_{r_+}^{r} \sqrt{g_{rr}} dr' 
  = \int_{r_+}^{r} \frac{dr'}{\sqrt{f(r')}}
  = - \log(\epsilon ) - 1  - (r + \log r) 
  + O\left( \frac{1}{r},~\epsilon \right),
\label{dD-perturbative}
\end{eqnarray}
where $f(r)$ is the metric function of the RN black hole 
defined by Eq.~\eqref{RNf}. 

Next we evaluate $L$ and $D$ in the DRN solutions.
In the DRN solution~(\ref{double_RN_metric})-(\ref{def-gamma-double-RN}),
we can choose $(z_1,z_2)=(-l,0)$ without loss of generality.
For this choice, the event horizon on the axis
is at $(\rho,z) = (0,\pm\sigma_2)$.
Then, the proper distance $L$ in the DRN solution is formally given by 
\begin{equation}
L = \int_{-\ell}^{- \sigma_2} \sqrt{F(0,z)} dz,
\label{dL-DRN}
\end{equation}
where $F(\rho,~z)$ is one of the metric functions of the DRN solution
defined by Eq.~\eqref{double_RN_f} 
and we take the upper end of the integral
at $z = -\ell $, which is sufficiently close to the singularity 
of the body 1. 
In the range of $-\ell < z < -\sigma_2$ on the symmetric axis,
from Eqs.~(\ref{bipolar}), we obtain  
\begin{align}\label{bipolar-coordinate-dL}
\begin{cases}
r_1 =   z + m_1 + \ell~, \quad \theta_1 = 0~, \cr
r_2 = - z + m_2~,     \quad \theta_2 = \pi~, 
\end{cases}
\end{align}
and then the metric function on the axis is given by
\begin{equation}
F(0,z) = \frac{[ (- z + m_2)(z + m_1 + \ell) -  e_1 e_2 ]^2}
{ [(z + \ell)^2 - \sigma_1^2]  [z^2 - \sigma_2^2]  }~.
\label{double-RN-f-dL}
\end{equation}
The direct integration of Eq.~\eqref{dL-DRN} gives a complicated combination
of incomplete elliptical integrals that is inconvenient for our analysis.
To avoid the complication, 
we divide the range of the integral $[-\ell,~-\sigma_2]$ into
$ [-\ell,~-\ell + A \epsilon^{3/2}]$ and 
$[-\ell + A \epsilon^{3/2},~-\sigma_2]$, where $A$ is an arbitrary
constant of $O(1)$. In the former range, we expand the term
$1/\sqrt{z^2-\sigma_2^2}$ in the integrand of Eq.~(\ref{dL-DRN}) with respect
to $z+\ell=O(\epsilon^{3/2})(\ll \sigma_2)$, keeping the term
$1 / \sqrt{ (z + \ell)^2 - \sigma_1^2 }$.
On the other hand, in the latter region, 
we expand $1 / \sqrt{ (z + \ell)^2 - \sigma_1^2 }$ with respect to
$\sigma_1$ while the factor $1 /\sqrt{z^2-\sigma_2^2}$ is kept unexpanded. 
Then, one can perform the integrals to obtain an approximate 
estimate of $L$ as
\begin{eqnarray}
L =
 \left( \int_{-\ell}^{-\ell + A \epsilon^{3/2}} 
 + \int_{-\ell + A \epsilon^{3/2}} ^{-\sigma_2} \right)  \sqrt{F(0,z)} dz 
  = {\rm{arccosh}}\sqrt{ \frac{a \tilde g} {g^2}} + O(\epsilon)~.
\label{dL-DRN-approxed-final}
\end{eqnarray}

The proper distance $D$ from the event horizon of the black hole
to the asymptotic region $z \gg 1$ along the symmetric axis 
is also formally given by 
\begin{equation}
D = \lim_{z \to \infty} \int_{\sigma_2}^{z} \sqrt{F(0,z')} dz'.
\label{dD-DRN}
\end{equation}
Here it should be noticed that, unlike the case of $L$,
we take the lower end of the integral at $z=\sigma_2$
because the integral is performed in the direction opposite 
to the body 1 (singularity).
In the range of the integral on the right hand side of
Eq.~(\ref{dD-DRN}), the bipolar coordinates are given by 
\begin{align}\label{bipolar-coordinate-dD}
\begin{cases}
r_1 = z + m_1 + \ell~, \quad \theta_1 = 0~, \cr
r_2 = z + m_2~,     \quad \theta_2 = 0~.
\end{cases}
\end{align}
Substituting Eqs.~\eqref{bipolar-coordinate-dD}
into Eq.~\eqref{double_RN_f}, the metric function is rewritten as 
\begin{equation}
\label{double-RN-f-dD}
F(0,z) = \frac{[ (z + m_2)(z + m_1 + \ell) -  (e_1 - 2 \gamma) e_2 ]^2}
{ [(z + \ell)^2 - \sigma_1^2]  [z^2 - \sigma_2^2]  }
= \frac{(z+m_2)^2}{z^2-\sigma_2^2} \left( 1 + O(\epsilon) \right).
\end{equation}
Then, the integral~\eqref{dD-DRN} with Eq.~\eqref{double-RN-f-dD} 
can be evaluated. 
%is given as
%\begin{equation}
%D =
%- {\log}\left(\frac{\epsilon}{2} \sqrt{\frac{\tilde g} {a}} \right)
%+ (z + \log z) 
%+ O \left(\frac{1}{z},~\epsilon \right).\
%\label{dD-DRN-approxed-temp}\
%\end{equation}
Since the coordinate $z$ is related to $r$ on the axis 
as $z=r-1+O(\epsilon)$, which can be easily verified  
by looking at, say, the lapse function, 
we finally obtain
\begin{equation}
\label{dD-DRN-approxed-final}
D =
- {\log}\left(\frac{\epsilon}{2} \sqrt{\frac{\tilde g} {a}} \right)
- 1  + (r + \log r) 
+ O \left(\frac{1}{r},~\epsilon \right).
\end{equation}
Comparing Eq.~\eqref{dD-perturbative} to
Eq.~\eqref{dD-DRN-approxed-final}
and Eq.~\eqref{dL-perturbative} to Eq.~\eqref{dL-DRN-approxed-final}, 
we establish relations between the two sets of parameters 
presented in Eqs.~(\ref{mapping}), and with the aid of 
the constraint equation~\eqref{discriminant} valid up to $O(\epsilon)$ 
we obtain $b$. 

%%%%%%%%%%%%%%%%%%%%%%%%%%%%%%%%%%%%%%%%%%%%%%%%%%%%%%%%%%%%%%%%%%%%%%%%
\section{ Evaluate the energy radiated to infinity from a moving charged 
particle}
\label{sec:flux}
%%%%%%%%%%%%%%%%%%%%%%%%%%%%%%%%%%%%%%%%%%%%%%%%%%%%%%%%%%%%%%%%%%%%%%%%
In this appendix we evaluate the energy $E_{\infty}$ emitted to infinity 
from a charged particle that falls into a RN black hole along a radial orbit.

\subsection{Energy flux formula}
First, we derive the effective energy flux 
generated from the scalar-type perturbation 
in the Kodama-Ishibashi (KI) formalism.
The energy flux carried by gravitational waves and 
that by electromagnetic waves are decoupled 
in the asymptotic region $r \to +\infty$,
and they are, respectively, given in terms of the perturbation of 
the metric $h_{\mu\nu}$ and that of the electro-magnetic field 
strength $f_{\mu\nu}$ as \cite{Misner:1974qy}
\begin{eqnarray}\label{radiation-infinity}
\dot E_{\infty}^{\rm{GW}} := - r^2 \int \frac{d\Omega}{4 \kappa}
{\partial h_{\alpha \beta}\over \partial t} 
{\partial h^{\alpha \beta}\over \partial r} ~,
\qquad
\dot E_{\infty}^{\rm{EM}}
= - r^2 \int d\Omega\,  {f}_{t \rho} {f}_{r}^{\,\rho}~,
\end{eqnarray}
in the transverse-traceless gauge defined by
$h^{\rho}_{\rho} = 0,~\nabla^{\nu} h_{\mu \nu} = 0$.  
Here, averaging over several wavelengths of radiation is assumed. 
We rewrite these expressions~\eqref{radiation-infinity} 
in terms of the master variables $\Phi_{\pm}$ in the KI formalism.
To do so, as an intermediate step we consider 
$\Phi$ and $\mathcal{A}$ that are related to
$\Phi_{\pm}$ at the level of coefficients of the spherical harmonics 
expansion by 
\begin{equation}\label{separate-M}
\Phi_{\pm} = a_{\pm} \Phi + b_{\pm} {\mathcal{A}},
\end{equation}
with the coefficients defined in Eqs.~\eqref{ab-coeff}, 
where the indices of spherical harmonics, 
${\rmx l}$ and ${\rmx m}$, are suppressed for notational simplicity. 
At the leading order in the limit $r \to +\infty$, 
the master variables $\Phi$ and $\mathcal{A}$ are, respectively, related to 
the perturbations of gravitational and electro-magnetic fields as
\begin{eqnarray}\
\label{boundary-Master-variables}
 h_{ij} := \sum_{\rmx l,\rmx m} {{\rmx l}({\rmx l}+1)} 
\, r\, \Phi\,
{\mathbb{S}}_{ij}, \qquad
 f_{ai} \approx \sum_{\rmx l,\rmx m} \sqrt{{\rmx l}({\rmx l}+1)} 
\, \epsilon_{a b} (D^{b} {\mathcal{A}})\, {\mathbb{S}}_i~, 
\end{eqnarray}
where $a, b$-indices run over $t, r$-coordinates, while 
$i, j$-indices over angular coordinates. The other components 
are of higher order in $1/r$. 
The totally anti-symmetric symbol $\epsilon_{a b}$ is defined 
so that $\epsilon_{tr}=1$. 
The harmonics are defined as 
\begin{equation}\label{scalar-harmonics}
\mathbb{S}     := Y_{\rmx l \rmx m}~, \quad
\mathbb{S}_{i}  := - \frac{1}{\sqrt{{\rmx l}({\rmx l}+1)} } 
\hat{D}_{i} Y_{\rmx l \rmx m}~, 
\quad
\mathbb{S}_{ij} := 
  \frac{1}{{\rmx l}({\rmx l}+1)}  \hat{D}_{i} \hat{D}_{j} Y_{\rmx l \rmx m} 
+ \frac{1}{2}   {\gamma}_{ij} Y_{\rmx l \rmx m}~, 
\end{equation}
where $Y_{\rmx l \rmx m}$ are the usual spherical harmonics on 
a unit sphere $S^2$ and $\hat{D}_i$ represents covariant differentiation 
with respect to the metric of $S^2$, $\gamma_{ij}$. 

Then, substituting Eqs.~\eqref{separate-M},
~\eqref{boundary-Master-variables} and 
~\eqref{scalar-harmonics} 
into Eq.~\eqref{radiation-infinity}, 
the total energy flux escaping to infinity is evaluated as
\begin{eqnarray}
\label{Flux-boundary-G}
\dot E_{ \infty } &\!\! = &
\!\! - \sum_{\rmx l,\rmx m} 4 \pi {{\rmx l}({\rmx l}+1)}  \left(
\frac{ \partial {\mathcal{A}} }{ \partial t } 
\frac{ \partial {\mathcal{\bar A} }}{\partial r}  
+
\frac{{({\rmx l}-1)({\rmx l}+2)} 
}{8 \kappa^2} 
 \frac{\partial \Phi}{\partial t} 
\frac{\partial \bar \Phi}{\partial r} 
\right) \cr
&\!\! =&\!\! - \sum_{\rmx l,\rmx m} 
{4\pi {{\rmx l}({\rmx l}+1)}  \over 9 \kappa^2 \nu(M + \nu)}
\left(
\frac{\partial {\Phi}_{+}}{\partial t}
\frac{\partial {\bar \Phi}_{+}}{\partial r}
+{{({\rmx l}-1)({\rmx l}+2)} \over 16}
\frac{\partial {\Phi}_{-}}{\partial t}
\frac{\partial {\bar \Phi}_{-}}{\partial r}
\right)~,
\end{eqnarray}
where quantities with ``$~\bar{~}~$'' represent 
the complex conjugations, and 
all terms suppressed by $1/r$ are neglected.
Note that this expression is valid 
for all modes including the exceptional mode with ${\rmx l}=1$. 
However, for the exceptional mode the second term in the parentheses 
on the last line vanishes. Hence, we find that $\Phi_-$ is irrelevant for 
this mode, which is consistent with the fact that physical degrees of 
freedom for the metric perturbation are absent for the exceptional mode.  
(See Ref.~\cite{Kodama_04} for further details.)

We would like to add a short remark 
about the expression for the energy flux~\eqref{Flux-boundary-G}. 
This is a special case of a more general expression for the energy flux 
defined as an integration over an arbitrary 2-surface of constant $r$:
\begin{equation}
\label{flux-G}
\dot E(r)
= - \sum_{\rmx l,\rmx m} 
{4\pi {{\rmx l}({\rmx l}+1)}f  \over 9 \kappa^2 \nu(M + \nu)}
\left(
\frac{\partial {\Phi}_{+}}{\partial t}
\frac{\partial {\bar \Phi}_{+}}{\partial r}
+{{({\rmx l}-1)({\rmx l}+2)} \over 16}
\frac{\partial {\Phi}_{-}}{\partial t}
\frac{\partial {\bar \Phi}_{-}}{\partial r}
\right)~,
\end{equation}
which is identical to Eq.~\eqref{Flux-boundary-G} in the 
limit $r \to {\infty}$.
Here, $f$ is the metric function of the background RN black hole 
given in Eq.~\eqref{RNf}. 
To verify that the above expression is conserved 
in the absence of source terms, 
we focus on the fact that the master equation~\eqref{MasterEq} 
is derived from the variational principle of the action 
\begin{equation}\label{scalar-action}
S := \int d^2 x \sqrt{-g_{(2)}} 
\left( -\frac{1}{2} g^{a b}_{(2)} 
\nabla^{(2)}_{a} \Phi_{\pm} \nabla^{(2)}_{b} \Phi_{\pm} 
- V_{\pm} \Phi_{\pm}^2\right)~,
\end{equation}
where $g^{a b}_{(2)}$ is the two dimensional metric composed of the 
$(t,r)$ components of the RN metric, $\nabla^{(2)}_{a}$ is 
the covariant differentiation with respect to $g^{a b}_{(2)}$
and $g_{(2)}$ is the determinant of this two dimensional metric.
The $t$ and $r$ components are labeled by the Latin indices, $a, b$.
%The first half of the Latin indices, $a, b$, run $t$ and $r$. 
Once the action~\eqref{scalar-action} is at hand, 
its variation with respect to $g^{ab}_{(2)}$ 
gives a symmetric tensor $T_{a b}$ that satisfies the 
conservation law $\nabla^{(2)}_{b} T_{a}^{~b} = 0$. 
Then, one can construct a conserved current 
$T_{ab} (\partial_t)^b$ associated with 
the background Killing field $(\partial_{t})$.
Thus, we find the conserved flux per unit coordinate time 
is proportional to 
\begin{equation}
T_{ab}n^a (\partial_t)^b \sqrt{g^{(2)}_{tt}}\propto 
   f{\partial\Phi_\pm\over \partial t} {\partial\Phi_\pm\over \partial
   r},
\end{equation} 
where $n_a$ is the unit vector normal to a surface of constant $r$.
Since the expression (\ref{flux-G}) is a linear combination 
of these conserved fluxes, it is guaranteed to be conserved. 
For any wave packet the time integral of this flux is independent 
of $r$, and agrees with the net energy flux evaluated at infinity. 
The net energy flux is also a quantity that must be independent 
of $r$. Therefore for any wave packet the time integral of this 
flux (\ref{flux-G}) evaluated at any radius gives the net energy flux, 
and we conclude that Eq.~(\ref{flux-G}) is 
indeed an expression for the conserved energy flux valid 
for any value of $r$.  

%%%%%%%%%%%%%%%%%%%%%%%%%%%%%%%%%%%%%%%%%%%%%%%%%%%%%%%%%%%%%%%%%%%%%%%%
%%===========================================================
\subsection{Evaluation of the energy flux from a radially falling particle}
%%%%%%%%%%%%%%%%%%%%%%%%%%%%%%%%%%%%%%%%%%%%%%%%%%%%%%%%%%%%%%%%%%%%%%%%
%%===========================================================

Now we turn to the evaluation of 
the coefficients ${\mathcal{X}}_{\pm}(\omega)$ in Eq.~\eqref{inhomo-sol}. 
The explicit form of these coefficients are, with 
$Y_{\rmx l 0}:= \sqrt{(2 {\rmx l} + 1) /4 \pi}$, expressed as
\begin{eqnarray}
{\mathcal{X}}_{\pm}(\omega) / Y_{\rmx l 0} &\!\! =\!\!&
%%%%%% S_{A} series %%%%%%%%
\label{Integrands}
               i \frac{8 q b_{\pm} }{ {\rmx l}( {\rmx l + 1} ) \omega } 
\int_{r_+}^{r_0} 
\left[ \frac{z f}{r^2 H}  
\Phi_{\pm}^{\rm{in}} e^{i \omega T} \right]dr \cr
%\label{Integrands02}
                &&-i \frac{2\sqrt{2} \kappa \mu Q  b_{\pm}}{\omega } 
\int_{r_+}^{r_0} 
\left[ \frac{1}{r^3 H} 
\left( f\frac{dT}{ds} \right)  
  \Phi_{\pm}^{\rm{in}} e^{i \omega T} \right]dr \cr
%%%%%% S_{A} series *partial integral*  %%%%%%%%
%\label{Integrands03-1}
                &&-i \frac{q b_{\pm}}{{\rmx l}( {\rmx l + 1} ) \omega } 
\int_{r_+}^{r_0} 
\left[{V_{\pm}\over f}
 \Phi_{\pm}^{\rm{in}} e^{i \omega T}  \right]dr \cr
%%% S_{\Phi} series %%%%%%%
\label{Integrands05-1}
                &&+i 
                \frac{\sqrt{2} \kappa qQ}{ {\rmx l}( {\rmx l + 1} ) \omega} 
\int_{r_+}^{r_0} \left[ \frac{P_{S1}}{r^3 H^2} 
a_{\pm} \Phi_{\pm}^{\rm{in}} e^{i \omega T} \right] dr \cr
%\label{Integrands05-2}
                 &&-i \frac{2\sqrt{2} \kappa qQ}{\omega } 
\int_{r_+}^{r_0} 
\left[ \frac{1}{r^3 H} a_{\pm} 
\Phi_{\pm}^{\rm{in}} e^{i \omega T} \right]dr \cr
%\label{Integrands06}
                 &&+i \frac{ \kappa^2 \mu}{\omega} 
\int_{r_+}^{r_0} 
\left[ \frac{P_{S3}}{r^2 H^2} 
\left(f \frac{dT}{ds} \right)a_{\pm} 
\Phi_{\pm}^{\rm{in}} e^{i \omega T} \right]dr \cr
%\label{Integrands07}
                 &&+ {2 \kappa^2 \mu} 
\int_{r_+}^{r_0} \left[ \frac{1}{r f H} 
\left( \frac{dR}{ds} \right)a_{\pm} 
\Phi_{\pm}^{\rm{in}} e^{i \omega T} \right]dr \cr
%%% S_{\Phi} series *partial integral*  J_{t} %%%%%%%
                 %% Cancel out to the term coming from \tilde{J}_r
%%% S_{\Phi} series *partial integral* S_{t}^{R}%%%%%%
%\label{Integrands08-1}
                 & &-i \frac{2 \kappa^2 \mu}{\omega} 
\int_{r_+}^{r_0} 
\left[ \frac{1}{r^2} \frac{d}{dr} 
\left( \frac{a_{\pm} r}{H} \Phi_{\pm}^{\rm{in}}\right) 
\left(f \frac{dT}{ds} \right)
 e^{i \omega T} \right]dr~, 
\end{eqnarray}
for both generic and exceptional modes. 
In the above expression, 
$V_{\pm}$ are the effective potentials 
of the scalar-type perturbation defined in Eq.~\eqref{potential-scalar}, and 
the functions $H,~P_{S1}$ and $P_{S3}$ are as defined 
in Eqs.~\eqref{source-aux}. The components 
$(t,r) = (T(s),R(s))$ specifies the trajectory of a charged particle
obtained by integrating 
Eqs.~(\ref{dR-ds}) and (\ref{dT-ds})
with the initial condition that the particle is at rest at the 
equilibrium position $r = r_0$ in the limit $t\to - \infty$.

Notice that the integrands in the above expression 
do not possess any divergences at $r=r_0$. 
By contrast, in the limit $r=r_+$ 
the integrands of the last two terms in Eq.~(\ref{Integrands}) diverge. 
To see this, recall that 
$f$ vanishes at $r=r_+$ and 
$dT / ds$ and $d\Phi_{\pm}^{\rm{in}}/dr$ 
behave like $\propto f^{-1}$ there, while the other functions 
including $dR/ds$ are regular.  
To ameliorate this singular behavior, we perform the 
integration by parts in the following manner. 
First, we rewrite $e^{i\omega T}$ in the second to last term 
by using the following identity 
\begin{equation}
  e^{i\omega T} = {1 \over i\omega}{dR \over dT} 
   \left( \frac{d e^{i\omega T}}{dr} \right).
\end{equation}
Then, using the relation $f(dT/ds)^2 - f^{-1}(dR/ds)^2 = 1$, 
which is simply the normalization condition of the four velocity, 
the contribution from the last two terms can be rewritten as 
\begin{eqnarray}
\label{lasttwoterms}
&&\mbox{( the last two terms in Eq.~\eqref{Integrands} )}
\nonumber \\ &&\qquad 
=
\frac{2 \kappa^2 \mu}{i\omega} 
\int_{r_+}^{r_0} 
\left[  
\frac{a_{\pm} }{r H} {dT \over ds}
 \Phi_{\pm}^{\rm{in}}
\left\{f -\left( {dT \over ds} \right)^{-2}
\right\}
 \left( \frac{ d e^{i\omega T} }{dr} \right) 
+
\frac{f}{r^2} {dT\over ds}
e^{i \omega T} 
\frac{d}{dr}  
\left(
\frac{a_{\pm} r}{H} 
 \Phi_{\pm}^{\rm{in}}
\right)
\right]dr
\cr&&\qquad =
\frac{2 \kappa^2 \mu}{i\omega} 
\int_{r_+}^{r_0} 
\left[  
{f \over r^2} {dT\over ds}
\frac{d}{dr}  
\left(
\frac{a_{\pm} r}{H} \Phi_{\pm}^{\rm{in}} e^{i \omega T} 
\right) 
-
\frac{a_{\pm} }{r H} 
\left({dT\over ds}\right)^{-1}\!\!\!
 \Phi_{\pm}^{\rm{in}}
{ \left( \frac{ d e^{i\omega T} }{dr} \right) } 
\right]dr
\cr&&\qquad = {-} 
\frac{2 \kappa^2 \mu}{i \omega} 
\int_{r_+}^{r_0} 
\left[  
\frac{a_{\pm} r}{H} \Phi_{\pm}^{\rm{in}} 
\frac{d}{dr}  
\left({ \frac{f}{r^2} }  {dT \over ds}
\right) 
-
\frac{d}{dr}  
\left\{
\frac{a_{\pm} f}{r H} 
\left(f {dT \over ds}\right)^{-1}\!\!\!
 \Phi_{\pm}^{\rm{in}}
\right\}
\right]e^{i \omega T} 
dr~,
\end{eqnarray}
where we performed integration by parts in the last equality and 
dropped 
the surface terms. The surface terms take the form of an 
infinitely oscillating function such as $e^{i \omega T}$ multiplied 
by a regular factor near the boundaries. 
Such terms can be safely neglected by introducing 
an infinitesimal damping factor as usual.  
In the final expression the integrand is free from divergence 
on both boundaries at $r=r_+$ and $r_0$.

%%%%%%%%%%%====================================
%%%==========-------------------------------------

%%%%%%%%%%%%%%%%%%%%%%%%%%%%%%%%%%%%%%%%
%%%
%%% bibliography
%%%%
%%%% extreme sugeee = cho sugeeee
%%%
%%%%%%%%%%%%%%%%%%%%%%%%%%%%%%%%%%%%%%%%%%%%

\end{document}